\documentclass[a4paper,11pt]{article}
\pdfoutput=1 

\usepackage{jcappub} 

\usepackage[utf8]{inputenc}	
\usepackage[T1]{fontenc} 		
\usepackage{lmodern}				
\usepackage{microtype}			

\usepackage{cleveref}
\usepackage{siunitx}
\usepackage[all]{nowidow}
\usepackage{hhline}

\usepackage[final]{changes} 	

\input{aas_macros.sty}

\renewcommand{\vec}[1]{\mbox{\boldmath$#1$}}

\def \d			{\mathop{}\!\mathrm{d}}							

\DeclareSIUnit \arcmin 	{arcmin}
\DeclareSIUnit \arcsec 	{arcsec}
\DeclareSIUnit \parsec 	{pc}
\DeclareSIUnit \eV 			{eV}
\DeclareSIUnit \Msun 		{M_\odot}
\DeclareSIUnit \cts			{cts}
\DeclareSIUnit \deg			{deg}
\DeclareSIUnit \ph 			{ph}

\definechangesauthor[name={brown}, color=brown]{br}
\definechangesauthor[name={cyan}, color=cyan]{cy}

\title{\boldmath New constraints on sterile neutrino dark matter from the Galactic Center}

\author[a,b]{R. Yunis,}
\author[c]{C.~R. Argüelles,}
\author[d]{Nick~E. Mavromatos,}
\author[e]{A. Moliné,}
\author[a,b,f]{A. Krut,}
\author[a,b,g]{J.~A. Rueda,}
\author[a,b,g]{R. Ruffini}

\affiliation[a]{ICRANet, Piazza della Repubblica 10, I--65122 Pescara, Italy}
\affiliation[b]{Dipartimento di Fisica and ICRA, Sapienza Università di Roma, P.le Aldo Moro 5, I--00185 Rome, Italy}

\affiliation[c]{Instituto de Astrof\'{i}sica de La Plata (CCT La Plata, CONICET, UNLP), Paseo del Bosque, B1900FWA La Plata, Argentina}

\affiliation[d]{Theoretical Particle Physics and Cosmology Group, Department of Physics, King's College London,
Strand WC2R 2LS, London, U.K.}

\affiliation[e]{Instituto de Astrofísica e Ciências do Espaço, Faculdade de Ciências da Universidade de Lisboa,
Edifício C8, Campo Grande, P-1749-016 Lisbon, Portugal}

\affiliation[f]{University of Nice-Sophia Antipolis, 28 Av. de Valrose, 06103 Nice Cedex 2, France}

\affiliation[g]{ICRANet-Rio, CBPF, Rua Dr.~Xavier Sigaud 150, Rio de Janeiro, RJ, 22290--180, Brazil}

\emailAdd{rafael.yunis@icranet.org}

\abstract{
We calculate \added[id=br]{the most stringent} constraints \added[id=br]{up to date} on the parameter space for sterile neutrino warm dark matter models possessing a radiative decay channel into X-rays. These constraints arise from the X-ray flux observations from the Galactic center \added[id=br]{(central parsec)}, taken by the XMM and NuSTAR missions. We compare the results obtained from using different dark matter density profiles for the Milky Way, such as NFW, Burkert or Einasto, to that produced by the Ruffini-Arg\"uelles-Rueda (RAR) fermionic model, which has the distinct feature of depending on the particle mass. We show that due to the novel core-halo morphology present in the RAR profile, the allowed particle mass window is narrowed down to $m_s\sim 10\text{ -- }\SI{15}{\kilo\eV}$, when analyzed within the $\nu$MSM sterile neutrino model. We further discuss on the possible effects in the sterile neutrino parameter-space bounds due to a self-interacting nature of the dark matter candidates.}

\begin{document}
\thispdfpagelabel{Title}
\noindent{\small preprint KCL-PH-TH/2018-48}\\
\maketitle
\flushbottom


\section{Introduction}
\label{sec:intro}

While the evidence for the existence of dark matter (DM)  is implied from astrophysical and cosmological observations of gravitational effects, a huge effort is still focused on the understanding of the nature of the particles that make up this unknown matter as well as its detection~\cite{Jungman:1995df, Bergstrom:2000pn, Munoz:2003gx, Bertone:2004pz}. Among the myriad of DM candidates proposed, a sterile neutrino with a mass in the $\si{\kilo\eV}$ range has been claimed as a viable candidate, falling in the category of \textit{warm} DM (WDM)~\cite{Berezhiani:1995, Chun:1999, Langacker:1998, Arkani-Hamed:2001, Abazajian:2003, Asaka:2005, Aliu:2005}. 

These claims seem to be revitalized given recent results about neutrino oscillations and several other physics phenomenons which are not predicted by the Standard Model (SM) and suggests new, unknown physics~\cite{Ahmad:2002, Ashie:2005, Araki:2005, Weinheimer:2003}. In particular, a \added[id=br]{minimal extension} of the SM, the so-called \textit{Neutrino Minimal Standard Model} ($\nu$MSM) introduces three sterile right handed neutrinos, where the lightest one might account for the presence of DM in the universe~\cite{Asaka:2005}. 

\begin{figure}
\centering
\includegraphics[width=0.9\hsize]{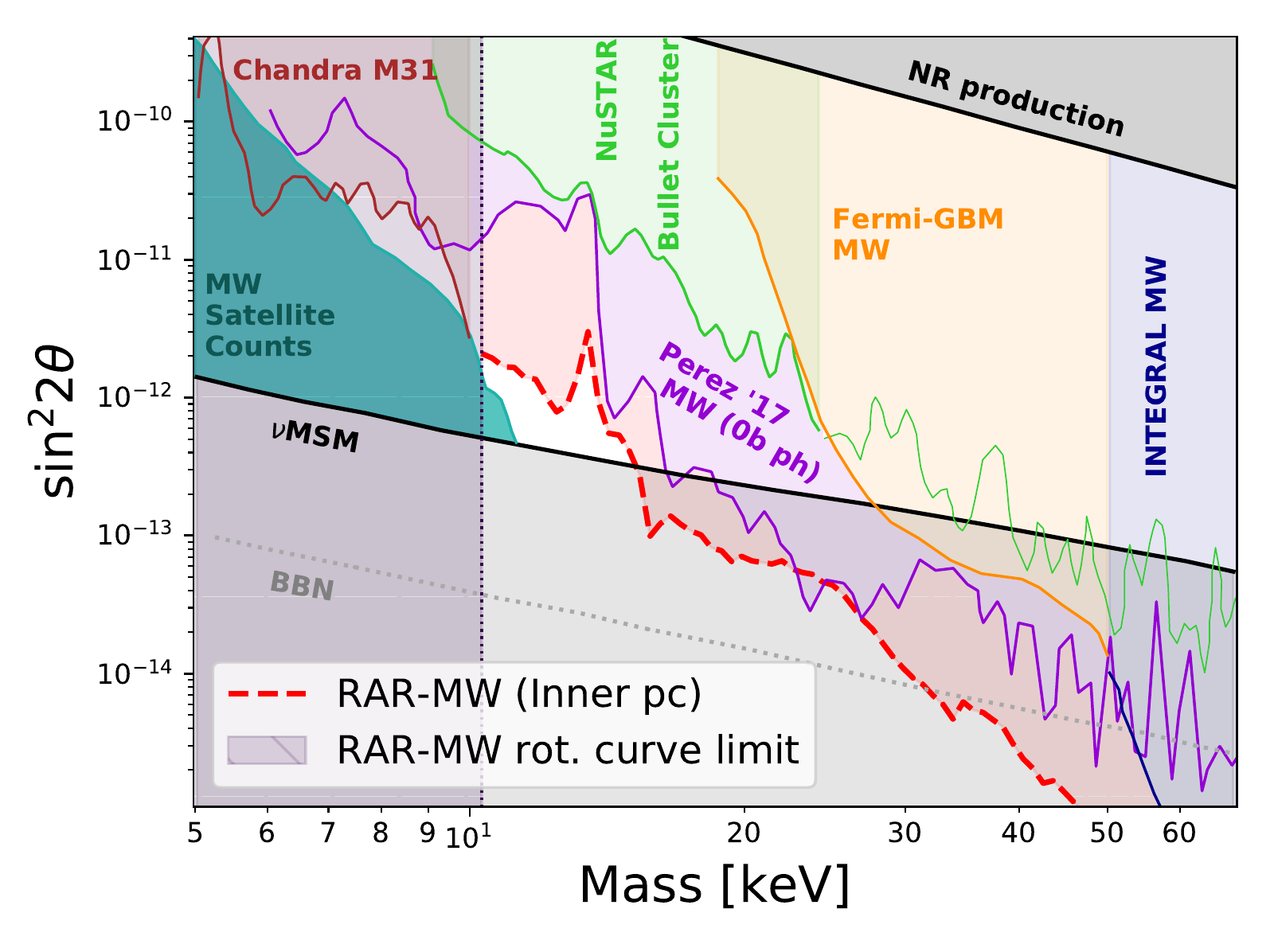}
\caption{Sterile neutrino parameter space limits obtained for GC observations using RAR profiles, compared to previous bounds. Red-dotted line and \SI{10.4}{\kilo\eV} vertical shaded region correspond to RAR limits given by X-ray bounds and MW rotation curve limits \cite{Arguelles2016} respectively. Upper and lower grey shaded regions correspond to production mechanism bounds: NRP under no lepton asymmetry, $\nu$MSM for maximal model produced asymmetry and maximal BBN allowed values \cite{Boyarsky2009,DodelsonWidrow94,AsakaLaineShaposhnikov}. Shaded blue region labels MW satellite count bounds \cite{BoyarskyLowerBound,HoriuchiHumphrey}. Other lines refer to several X-ray bounds \cite{HoriuchiHumphrey,BoyarskyMalyshev,NgHoriuchi,RiemerSorensen,NeronovMalyshev} including 0-bounce photon analysis \cite{Perez2017}, discussed further in \cref{sec:comparison}.}
\label{fig:Grafico_RAR_vs_P2017}
\end{figure}

From the gravitational evidence about the existence of DM, we can infer that the DM particle  must be stable on cosmological time scales. \added[id=br]{Nevertheless}, huge amounts of DM particles can decay even for such extremely long lifetimes and the decay signals
may be in the observable range to be detectable
~\cite{Eichler:1989, 2009JCAP...01..043N, Arvanitaki, Hamaguchi, Buchmueller, Chen, Ibarra3:2009, PerezPileviez, Shirai, CHEN200971, Ibarra:2009, Ibarra:2008, Ibarra2:2008, Ishiwata:2008, Pospelov, Jung-Bae, CHEUNG2009293, FUKUOKA2009401, Ibarra:2010, Covi:2009, Hisano:2009, Liu:2009, HISANO2009101, Buckley:2010, Yin:2009, Bumseok, Demir:2010, Ibarra4:2009}. 
Within the framework of the $\nu$MSM, the hypothetical sterile neutrino \added[id=br]{is an example of decaying DM} with a lifetime several times greater than the age of the Universe~\cite{Asaka:2005}. 
Viability for sterile neutrinos as to constitute the entirety of cosmological DM \added[id=cy]{requires} low enough mixing with the standard sector, measured by \added[id=cy]{the \textit{mixing angle}}
\begin{equation}
	\label{mixingangle}
	\theta^2=\sum_{\alpha=e,\mu,\tau}(v^2 F_{\alpha,1})/m_{s}^2\, ,
\end{equation}
with $v^2$ the Higgs boson v.e.v., $F_{\alpha,1}$ the Yukawa couplings for the right-chiral neutrinos and $m_{s}$ the sterile neutrino mass. Indeed, \added[id=cy]{it is enough for $\theta^2$ to fall below $\num{2.5E-13} (\si{\mega\eV}/m_{s})^5$} as to have a lifetime longer than the age of the universe for tree-level decays~\cite{Dolgov02}. \added[id=br]{More stringent observational bounds, as the ones arising from the diffuse X/$\gamma$-ray background, assure the fulfillment of this life/time condition~\cite{Dolgov02}, with corresponding bounds falling just outside the upper-right end limits of the parameter space coverage, e.g. \cref{fig:Grafico_RAR_vs_P2017}}.
Both the sterile neutrino mass $m_s$, and the mixing angle between active and sterile neutrinos $\theta$, constitute the parameter space for $\nu$MSM  regarding DM. Their decay open the possibility of indirectly detecting DM via the identification of such interactions products as photons or neutrinos. Indeed, the sterile neutrinos would have a subdominant radiative decay channel into a photon and a light (mostly active) neutrinos~\cite{Pal:1982, Abazajian:2007}. An important clue for searching for the decay of a sterile neutrino candidate may be coming from the X-ray observation of DM-dominated objects, such as galaxies and clusters of galaxies. The region of the galaxy with the highest DM density is the Milky Way  center and constitutes the classical target for DM searches (see e.g.~\cite{2010pdmo.book.....B} for a compilation of works). The Galactic center (GC) region has been extensively studied in the realm of sterile neutrino DM decays using observations of several X-ray satellites such as Suzaku, Chandra, XMM-Newton (see \cite{2017JCAP...01..025A} for a recent review), as well as from the NuSTAR mission \cite{Perez2017}. In this paper, we focus on the possibility of detecting the X-ray signal produced in the GC due sterile neutrino decays, and compare among different hosting DM halo profiles.

The several works cited just above were developed to constraint the $\nu$MSM parameter space using X-ray observations from both galactic and extragalactic objects. Such limits come as complementary to the ones imposed by production mechanisms of DM in the early universe, such as non-resonant (Dodelson - Widrow) production \cite{DodelsonWidrow94,AsakaLaineShaposhnikov,2017JCAP...01..025A} and resonant production \cite{LaineShaposhnikov,ShiFuller,AsakaLaineShaposhnikov}. Other limits to the $\nu$MSM model include phase space distribution bounds, \added[id=br]{as well as bounds from local group galaxy counts, which exclude masses below several $\si{\kilo\eV}$} \cite{BoyarskyLowerBound,HoriuchiHumphrey,Schneider}. While X-ray bounds provide upper limits to the mixing angle $\theta$ (between dark and active sector) as a function of particle mass $m_{s}$.

The intensity of the DM decay flux expected from a individual halo depends mainly on the DM density distribution inside it. N-body simulations within the $\Lambda$CDM paradigm, seem to point towards a single universal description for the the DM halo density profiles, and different parameterizations had been obtained in the literature~\cite{Navarro:1995iw,Navarro:1996gj,MNR:MNR3039,Klypin:2000hk,Navarro:2003ew}. However, in this work we are interested in sterile neutrinos pertaining to the WDM paradigm with particles decoupling while still relativistic, implying a subsequent free-streaming which damp primordial density fluctuations below a cutoff scale sensitive to the particle mass (see e.g.~\cite{Boyarsky2009}). For keV-ish particles, such a damping imply a suppression in the power spectrum which goes from $\sim 10\%$ at Mpc scales when compared with the $\Lambda$CDM one, and becomes stronger on scales below $10^2$~kpc (see~\cite{2012PDU.....1..136B} for current constraints). One of the main consequences of such a suppression is the difference in morphology of the DM density profiles: while CDM halos (and sub-halos) are cuspy through the center, the WDM ones tend to form cored inner halos for low enough particle masses \textit{below} keV (see e.g.~\cite{2012MNRAS.424..684S}). Nevertheless, in the case of few to several keV, recent high resolution N-body simulations \cite{2017MNRAS.464.4520B} developed to understand the small-scale structure differences between CDM and WDM cosmologies, show that the density profiles in WDM halos with masses $M\gtrsim 10^{10} M_{\odot}$\footnote{Less massive WDM halos do start to show systematically lower concentrations respect to CDM ones \cite{2017MNRAS.464.4520B}.} are indistinguishable from their CDM counterparts, and well fitted by Einasto profiles~\cite{Navarro:2003ew}, or even by NFW \cite{2012MNRAS.424..684S} (though for much larger halo masses in the later). Therefore, in this paper we will compare  the Navarro-Frenk-White (NFW) profile~\cite{Navarro:1995iw, Navarro:1996gj}, the Einasto (EIN) profile~\cite{Einasto2006,Navarro:2003ew} and the Burkert (BUR) profile~\cite{Burkert:1995}, with the recently proposed Ruffini-Arg\"uelles-Rueda (RAR) profile~\cite{RAR1,Arguelles2016} in order to bracket the theoretical uncertainty in the limits from the modeling of the expected DM decay flux of sterile neutrino WDM.

At this point it is important to emphasize that the RAR model may imply a more self-consistent approach to deal with keV WDM particles instead of the use of NFW (or Einasto, etc), given (i) the limitations of N-body numerical resolution below a given radial scale (usually below $\sim 10$~kpc); and (ii) the fact that for dwarf galaxies the density profile fits clearly deviates from NFW (see e.g.~\cite{2017MNRAS.464.4520B}). Besides, the RAR profile presents distinct features that make up qualitative differences in the analysis, such as an explicit particle mass dependence in the profile itself, and a novel high-density (sub-pc) core density working as alternative to the central BH in addition to the outer halo~\cite{Arguelles2016} (see \cref{sec:RAR}). We further estimate the upper limits on the sterile neutrino parameter space ($\theta$, $m_{s}$), comparing the X-ray flux observations from the GC against the (DM halo model dependent) theoretical expected one. In particular, we focus on searches with NuSTAR and XMM-Newton satellites which have provided accurate observation of diffuse X-ray emission within the central region of the Galaxy.
Recently, such approaches have been undertaken in the literature using NuSTAR data \cite{Perez2017} using corona-like regions around the GC, where diffuse photons are included but excluding the very center (where many individual bright X-ray sources can be identified). Interestingly, when diffuse emission from the very few parsecs around SgrA* is included in the analysis, \added[id=br]{the RAR density spikes (i.e. quantum cores) which are sensitive to the particle mass, imply the more stringent} constraints in the (up to date) free parameter space. \added[id=br]{The main result from such an analysis accounting for RAR profiles, is that the $\nu$MSM free parameter window is significantly reduced or nearly ruled-out. Nevertheless}, we point out in \cref{sec:interactingDM} that novel generation mechanisms \added[id=br]{for sterile neutrinos} acting prior to the $\nu_{s} \rightarrow \nu + \gamma$ process, can relax considerably the $\nu$MSM free parameter space constraints.

In \cref{sec:decay} we describe the calculation of the expected DM decay flux. We summarize the relevant ingredients to compute it and the choice of parameters of the $\nu$MSM. We consider several parameterizations for the DM density profile and discuss its critical role in the expected DM decay flux. Using the X-ray observations from the GC, in \cref{sec:bounds} we obtain the upper limits on the sterile neutrino mixing angles as a function of particle mass. In addition,  we describe and compare bounds to the $\nu$MSM parameter space from other similar works on the field. In \cref{sec:interactingDM}, we discuss the effects in the relaxation of these bounds due to hidden dark sector interactions recently proposed by some of the authors of this work. Finally, in \cref{sec:concls} we draw our conclusions.

\section{Sterile neutrino decay: X-ray flux}
\label{sec:decay}

The radiative sterile neutrino decay channel to a photon  and an active neutrino produces a spectral line in the X-ray. The decay width is proportional to $m_{s}^{5}\sin^2(2\theta)$, where $\theta$ is typically a small quantity \added[id=br]{below the electroweak scale, as arising from Cosmic X-ray Background (CXB) constraints}~\cite{Abazajian:2007}. The expected energy flux observed from the decay of a massive neutrino will depend on both the distance and distribution of DM across the field of view of the detector, as well as on the parameter space ($\theta$, $m_{s}$).

This decay channel is due to mixing between active and sterile sectors under the $\nu$MSM model. The interaction arises due to mass mixing thanks to the addition of a Majorana mass term for the sterile neutrinos, plus a Yukawa Higgs-portal term as:

\begin{equation}\label{yuk_s2}
{\mathcal L}_{\rm Yuk} = F_{\alpha I} \, {\bar \ell}_\alpha \, N_{R\, I} \phi^c   + {\rm h.c.}~,  \quad I=1,2,3
\end{equation}
where $\ell_\alpha$ are the lepton doublets of the Standard Model (SM), $\alpha = e, \mu, \tau$, $F_{\alpha I}$ are the appropriate Yukawa couplings which relates to the mixing angle $\theta$ via \cref{mixingangle} above, and $\phi^c$ is the SM conjugate Higgs field, \emph{i.e}. $\phi^c = i \tau_2 \phi^\star$ (with $\tau_2 $ the $2 \times 2$ Pauli matrix), with $N_{R\, I}$ the three sterile neutrino fields. Further details will be explored in \cref{sec:interactingDM}.


If $x$ denotes the linear coordinate along the line of sight (l.o.s.) and $d\Omega$ the solid angle element of the detector field of view, the differential photon flux from a volume element in the galaxy\footnote{
	Ignoring general relativistic effects on proper volume integrals.
} $x^{2} dx d\Omega$,  reaching a unit effective area of the detector is proportional to the DM density profile $\rho$, and given by
\begin{equation}
\d f=x^{2}\d x \d\Omega\frac{\Gamma }{m_{s}}\frac{\rho(r)}{4\pi x} ~,
\end{equation}
since each volume element contains $\rho(r)/m_{s}$  sterile neutrinos. In the $\nu$MSM, the decay width $\Gamma$, is defined as~\cite{Pal:1982, BARGER1995365}
\begin{equation}
\Gamma = \frac{9}{1024}\frac{\alpha G_{\rm F}^{2}}{\pi^{4}}m_{s}^5\left | \Theta \right |^{2} ~
\label{eq:decay_rate}
\end{equation}
with $\alpha$  the fine-structure constant, $\left | \Theta \right |^{2} = \sin^2(2\theta)$ and $G_{\rm F}$ the Fermi constant. The average flux observed in a solid angle is then found by integrating over the $\rho$  along the line of sight connecting the detector and the GC and the solid angle,
\begin{equation}
F=\frac{\Gamma}{4 \pi m_{s}}\int_{\Omega_{\rm l.o.s.}}  \d\Omega \int \rho (r(x,\Omega))\, \d x ~.
\label{eq:exp_flux}
\end{equation}
This expression can be cast as
\begin{equation}
F=\frac{\Gamma}{4 \pi m_{s}} S_{\rm DM} ~,
\label{eq:S_factor_definition}
\end{equation}
where the $S_{\rm DM}$ factor contains both the features of the DM profile of interest and the observation details such as the location of the observed region itself. The remaining factor depend exclusively on the particle physics parameters and $\nu$MSM specific decay rates. \added[id=br]{By asking $F^{\rm obs}_{\rm max}\geq F$ with $F^{\rm obs}_{\rm max}$ the maximum observed X-ray flux (see \cref{sec:signal})}, it is possible to place upper limits on the sterile neutrino mixing angle as a function of particle mass, as shown in the picture below. 
Two main Milky Way observations/DM profile pairs are shown: expected fluxes corresponding to flat core NFW profile (motivated by observations, see~\cite{Perez2017}) are compared to NuSTAR observation for diffuse light $\sim$~100~pc around the GC, as reviewed in~\cite{Perez2017}, placing an upper limit on the sterile neutrino mixing angle. In this work, observations of the very inner $\sim$~few pc of the Milky Way were used and compared against RAR profile. An estimate for this upper bound is obtained, resulting in the more stringent limits within the $\nu$MSM parameter space \added[id=br]{up to date}.

\subsection{DM halo density profiles}
\label{sec:profiles}

DM density profiles for galaxy halos have been reviewed by many authors and is still a topic of discussion. As mentioned above,  since the expected photon flux from DM decays is proportional to the DM distribution inside the halo, the density profile plays a critical role in DM searches. 
One of the most commonly used parametrization is the NFW profile~\cite{Navarro:1995iw, Navarro:1996gj}
\begin{equation}
\rho_{\rm NFW}(r)=\frac{\rho_{\rm s}}{\left(r/r_{\rm s}\right)\,\left(1+\left(r/r_{\rm s}\right)\right)^2} ~,
\end{equation}
where $r_{s}$ is the scale radius and $\rho_{s}$ is the dark matter density at the scale radius. In this work, we consider the local DM density $\rho_{local}=\rho_{NFW}(r=r_{\odot}=\SI{8}{\kilo\parsec})= \SI{0.4}{\giga\eV/\centi\metre^3}$ and $r_{s}=\SI{21}{\kilo\parsec}$, which are compatible with the preferred parameters for the MW halo reviewed in \cite{ReadLocalDM}.

In order to study the impact of the density profile choice in the calculation of the $S$ factor, we consider other  alternative functional forms as the Einasto profile \cite{Einasto2006},
\begin{equation}
\rho_{\rm EIN}=\rho_{\rm s}\, {\rm exp} \left\lbrace -\frac{2}{\alpha} \left[ \left(\frac{r}{r_{\rm s}}  \right)^{\alpha}-1 \right] \right\rbrace ~,
\end{equation}
and the Burkert profile \cite{Burkert:1995},
\begin{equation}
\rho_{\rm BUR}(r)=\frac{\rho_{\rm s}}{\left(1+r/r_{\rm s}\right)^{\alpha}\,\left(1+r/r_{\rm s}\right)^{\beta}} ~,
\end{equation}
In this case, our choice is $r_{s}=\SI{21}{\kilo\parsec}$, $\alpha=0.17$ for the Einasto profile and $r_{s}=\SI{6}{\kilo\parsec}$, $\alpha=1$ and  $\beta=2$ for the Burkert profile. We also consider for both parametrization the same local DM density as described above.

While DM only simulations of the Milky Way favor profiles with density slopes similar to NFW at small radius, the scenario changes with the addition of baryons. Reference \cite{Calore}, which considered simulated galaxies with baryons that best fit the Milky Way data showed that the density slope is steeper for $1.5\text{ -- }\SI{6}{\kilo\parsec}$, and shallower below $\SI{1.5}{\kilo\parsec}$ compared to NFW.
A conservative approach to this data is considering a density profile identical to NFW, but with constant density below the $\SI{1.5}{\kilo\parsec}$ range. We denote this profile as coreNFW. \added[id=cy]{In addition, as complementary to those based on standard cosmological simulations it is possible to obtain the distribution of DM in a halo through a semi-analytical approach, which  has  the  chance  to  include  more  rich  physical  ingredients,  such  as  quantum  statistics,  thermodynamics, and gravity as we will briefly describe in the next section.}


\subsection{RAR profile}
\label{sec:RAR}

The recently proposed RAR profile \cite{RAR1} is based on a self gravitating system of massive fermions at finite temperatures. Its formulation and preferred parameters for the MW halo \added[id=br]{morphology can be seen in \cite{Arguelles2016}, as well its implications on MW rotation curve observables from the center to periphery}. These profiles carry distinct features that differentiate them from the ones above: first of all have an explicit dependence on fermion mass, a parameter which only appears in the decay rate factor for previous profiles, thus making the $S_{\rm DM}$ factor not completely independent of the decay rate particle model. Also, these types of profiles harbor \added[id=br]{highly} dense compact quantum cores, which are of interest \added[id=br]{when including the innermost S-star cluster around SgrA*, given that such cores are shown to work as an alternative to the central BH~\cite{Arguelles2016}}. 

\begin{figure}
	\centering
	\includegraphics[width=0.77\hsize,clip]{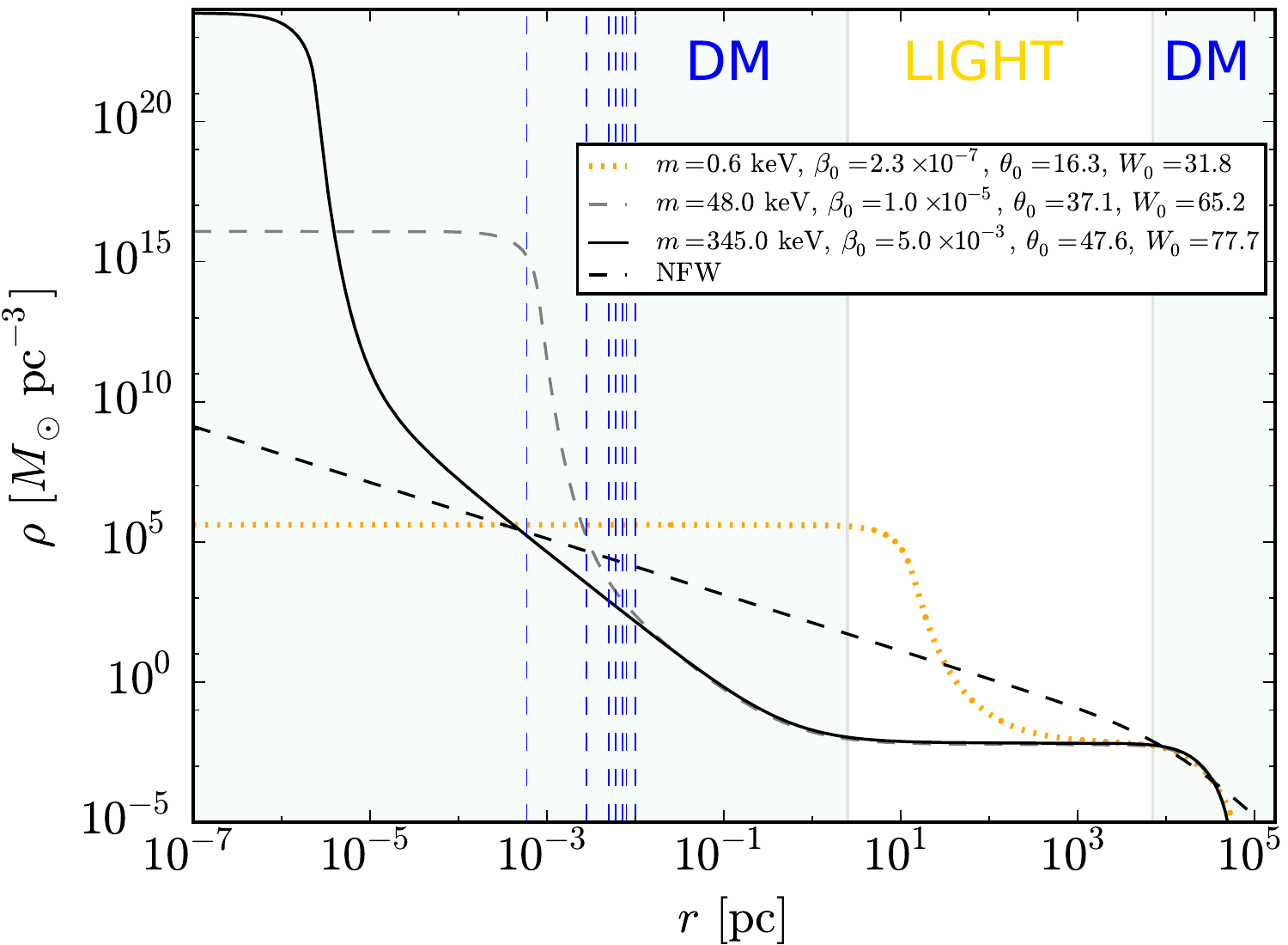}\\
	\includegraphics[width=0.77\hsize,clip]{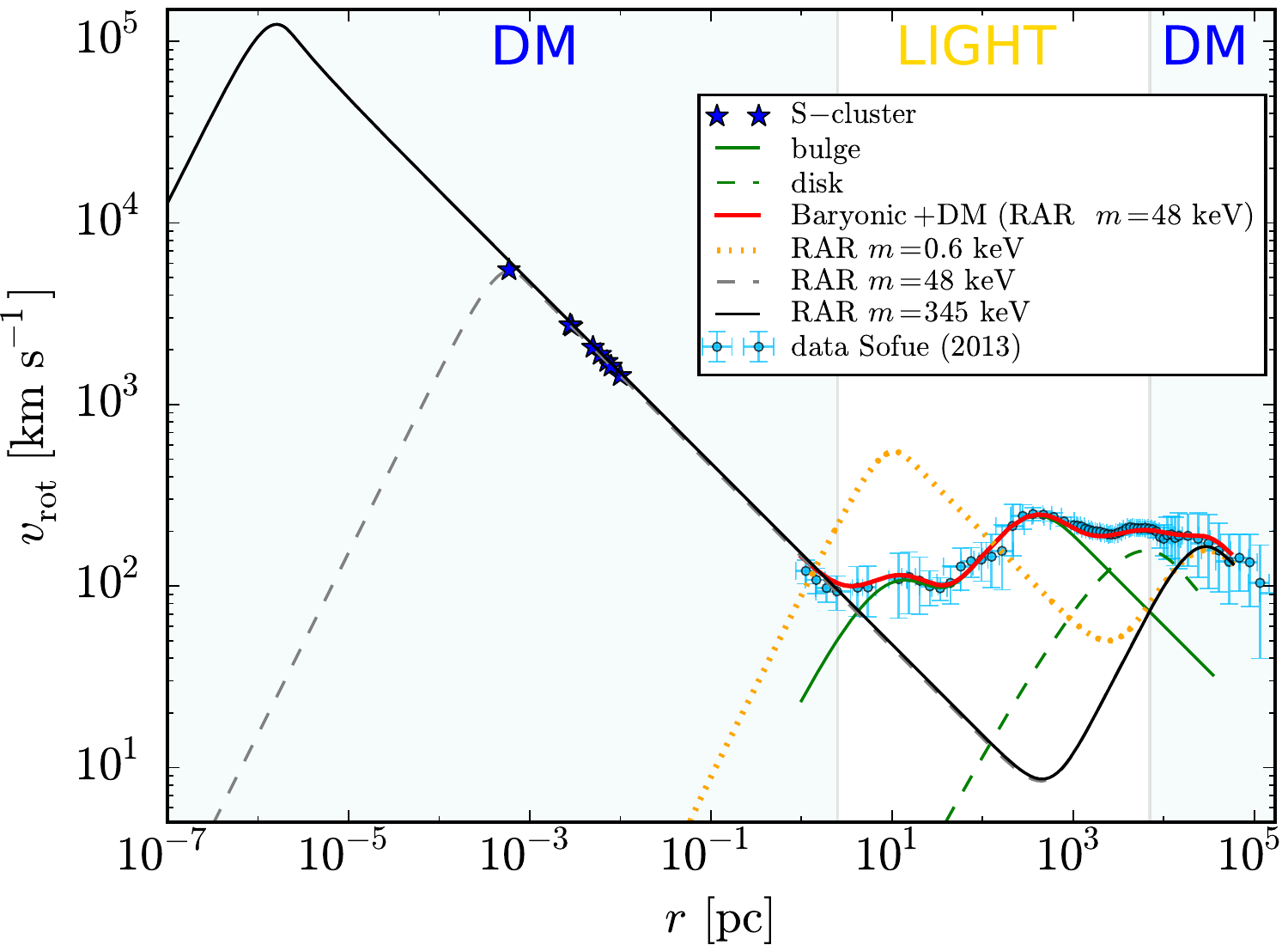}
	\caption{(Color online) Theoretical density profiles and rotation curves from $\SI{E-7}{\parsec}$ all the way to $\SI{E5}{\parsec}$, for three representative fermion masses in the $mc^2\sim\si{\kilo\eV}$ region: $\SI{0.6}{\kilo\eV}$ (dotted yellow curve), $\SI{48}{\kilo\eV}$ (long-dashed gray curve) and $\SI{345}{\kilo\eV}$ (solid black curve), with corresponding free RAR model parameters: ($\beta_0$,$\theta_0$,$W_0$) reading for central temperature, degeneracy, and escape parameters respectively. The RAR solutions for $mc^2 = 10.4\text{ -- }\SI{345}{\kilo\eV}$ are in agreement with the observed rotation curve from pc scales and on (i.e. Sofue-data). For the case of $mc^2=\SI{48}{\kilo\eV}$, we include the total rotation curve (red thick curve) including the total baryonic (bulge + disk) component. The dashed blue lines in upper panel indicate the position of the S-cluster stars \protect\citep{2009ApJ...707L.114G}. The NFW density profile is shown for sake of comparison as obtained in \cite{2013PASJ...65..118S} (dashed black curve). Fig. taken from \cite{Arguelles2016}.}
	\label{fig:RAR-MW}
\end{figure}

There are no closed analytical forms for these profiles, but can be integrated from a set of ordinary nonlinear integro-differential equations \added[id=br]{for different particle masses, with corresponding other free RAR model parameters such as central temperature, degree of central degeneracy and particle escape energy}. \added[id=br]{Interestingly, in \cite{Arguelles2016} it was explicitly shown that there is a unique particle mass range $10.4\text{ -- }\SI{345}{\kilo\eV}$ such that the RAR continuous solutions are in agreement with the full MW rotation curve (i.e. from $\sim 1\text{ -- }\SI{5E4}{\parsec}$), without spoiling the baryonic contribution which dominates at bulge/inner-disk scales (see \cref{fig:RAR-MW}). Moreover, such fermion masses obtained only from rotation curve observables, naturally find sterile neutrino (primordial) candidates from early Universe~\cite{Asaka:2005,2015PhRvD..92j3509P}, in agreement with other recent cosmological constraints such Ly-$\alpha$ forest~\cite{2009JCAP...05..012B,2017JCAP...06..047Y}, CMB observations and small scale structure (see~\cite{2017JCAP...01..025A} for a review), among others. The lower end of the mass range ($\SI{10.4}{\kilo\eV}$) is obtained under the condition that the total rotation curve (DM+baryons) remains consistent with MW observations, while the upper mass range ($\SI{345}{\kilo\eV}$) is the limiting mass before the whole configurations become unstable~\cite{Arguelles2016}}. The RAR profiles firmly excludes particle masses below the $\si{\kilo\eV}$ range, given that the central (too extended) cores overshoot the observed rotation curve in the bulge region (see \cref{fig:RAR-MW}). Result which is in line with totally independent phase-space (Tremaine$\&$Gunn-like) bounds.

\section{Signal analysis}
\label{sec:signal}

So far we have introduced all the ingredients in order to perform the analytical calculation of the DM decay flux.
By comparing the expected flux defined in \cref{eq:S_factor_definition} with the observed one \added[id=br]{(i.e. such that $F^{\rm obs}_{\rm max}\geq F$, see Sec. below)} it is possible to place upper limits on the sterile neutrino mixing angles as a function of particle mass. In this way, an astrophysical region must be selected, \added[id=br]{task which is presented in this section from NuSTAR and XMM-Newton observations}. The more stringent limits to the $\theta$ parameter will come from regions with a low maximal observed flux (i.e. few observed photons) and a high $S$ factor; then a high expected theoretical flux (i.e. high expected photons), as \added[id=br]{evidenced} in \cref{eq:S_factor_definition}. A suitable selection for the observation region must then fulfill such both conditions.

\subsection{Galactic Center}
\label{sec:SA_GC}

The \added[id=br]{main conclusions of this work (i.e. most stringent constraints to sterile neutrino parameter space) arise when considering the observations (photon flux) coming from} the innermost parsecs of the Galaxy. This observation is centred around G369.95-0.04, which is a Pulsar Wind Nebula candidate \added[id=br]{located at about} less than $\SI{9}{\arcsec}$ away from SgrA*, which we identify as the geometrical center of all DM density profiles \added[id=br]{here adopted}. The observation spans a circular region of $\SI{40}{\arcsec}$ around the centroid of G369.95-0.04. \added[id=br]{The observational data with corresponding spectra} has been obtained by the NuSTAR instrument, \added[id=br]{as presented} in \cite{Mori2015}.


\begin{figure}
\centering
\includegraphics[width=0.5\hsize]{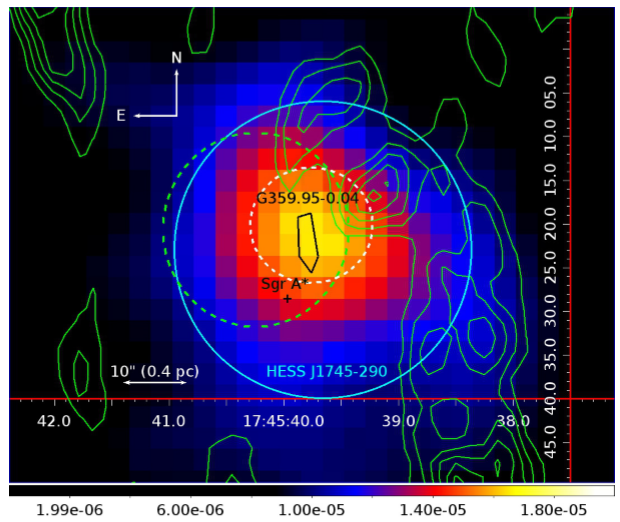}
\caption{NuSTAR $20\text{ -- }\SI{40}{\kilo\eV}$ band image zoomed in the central 30 region overlaid with SgrA* (black cross), the centroid of the TeV source HESS $\mathrm{J1745} - \mathrm{290}$ (cyan circle), PWN candidate $\mathrm{G359.95} - \mathrm{0.04}$ (black polygon) and circumnuclear disk (green contours). Reproduced from \cite{Mori2015} for didactical purposes. }
\label{fig:G359.95-0.04}
\end{figure}

The spectral analysis of this region shows a rich variety of X-ray sources in the $2\text{ -- }\SI{40}{\kilo\eV}$ band, according to \cite{Mori2015}. Such features include SgrA*, G359.95-0.04, SgrA East, stellar winds, element lines and the CHXE,\footnote{
	Central Hard X-ray Emission. According to the detailed spectral study of two nearby intermediate polars and the CHXE by \cite{Hailey:2016}, the CHXE emission is likely an unresolved population of massive magnetic cataclismic variables (CVs).
} among others.

This is an observation area filled with X-ray sources, and is expected to have a high observed photon flux. Indeed, \added[id=br]{we have performed such a $F^{\rm obs}_{\rm max}$} analysis for a set of energies in the range $2\text{ -- }\SI{50}{\kilo\eV}$. The results can be seen in \cref{fig:Grafico_maxflux}. We can see a maximal observed photon flux of about $\sim \SI{E-2}{\cts\ \second^{-1}\metre^{-2}}$ for a region of about $\SI{4E-4}{\deg^2}$ (total solid angle area of a $\SI{40}{\arcsec}$ circle around G359.95-0.04). \added[id=br]{Interestingly}, some DM profiles show a significant density increase at the inner parsecs of the galaxy, which leads to a boost in $S_{\rm DM}$ factor for those regions. RAR profiles are the best example, as the inner compact core accounts for most of the $S_{\rm DM}$ factor contributions, several orders of magnitude above other profiles \added[id=br]{for the same central area} (as seen in following sections).

\subsubsection*{Maximal line flux}

In order to successfully obtain limits on \added[id=br]{the sterile neutrino DM parameter space} from observations, it is necessary to determine a maximum X-ray flux that could have possibly been originated from DM decay.

Null-detection hypothesis has been tested by \cite{Perez2017} for the GC region within $\SI{E2}{\parsec}$ from SgrA*.\footnote{
	Namely, using the 0-bounce photon analysis for the GC data \added[id=br]{as also} considered in this paper \added[id=br]{for comparison purposes}.
} As this condition is independent from DM halo modeling and relative instrumental errors are unchanged, we assume the hypothesis to hold for \added[id=br]{the central few pc} region as well. \added[id=br]{Moreover}, given the fact that a best fit \added[id=br]{to the total observed Flux} including only astrophysical sources has been obtained \cite{Mori2015}, using the same instrument and within expected error bounds. \added[id=br]{A detailed analysis based on the modeling of all the known sources within the central pc, leading (or not) to the explicit null-detection conclusion will appear somewhere.}

Currently, observed spectra for the diffuse emission of the GC on the central $\sim$few~pc show features that could all be accounted for with astrophysical sources. Then, decays in the X-ray band could only have fluxes that fall within the statistical uncertainty for the measured spectra. Sterile neutrino decay modes can be approximated as monochromatic within the energy definition of current instrument, so the expected shape of a DM decay line in an observed spectrum would be a Gaussian peak no larger than the 3$\sigma$ confidence interval on the spectral fit, and with a width determined by the instrument’s energy definition.

Following the criterion used in \cite{HowToFindWDM}, we approximate the 3$\sigma$ confidence level for a spectra at a given energy by calculating the confidence intervals of a power law approximation centered around that given energy.

\begin{figure}
\centering
\includegraphics[width=0.65\hsize]{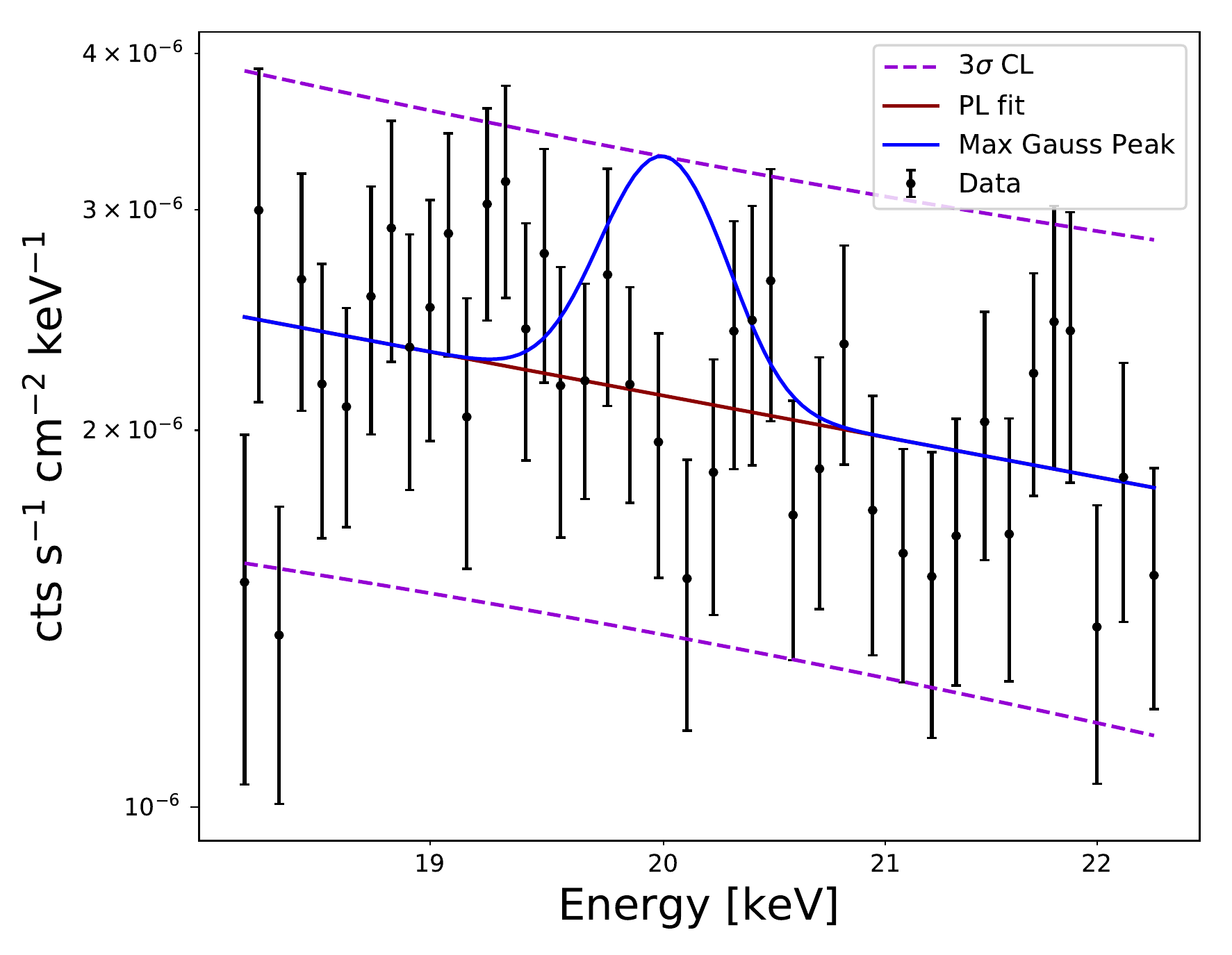}
\caption{An example of a maximal observed flux analysis for energy $\SI{20}{\kilo\eV}$. We have used the $\SI{40}{\arcsec}$ circular GC region diffuse emission spectra from \cite{Mori2015}. The black dots show an example data set, solid lines the power law (PL) and maximal Gaussian peak fits (red and blue respectively) with 3$\sigma$ confidence bounds for the PL fit on dashed lines. }
\label{fig:Grafico_ej_maxflux}
\end{figure}

Then, a maximal DM decay line would be seen in the spectra as a Gaussian peak, with a flux not exceeding the spectra plus its 3$\sigma$ error and as wide as the instrument's resolution for such energy. This line would give an upper limit for the total DM decay flux, calculated as the integral of such maximal Gaussian peak. As an example, we show one of this maximal peaks for a given energy in \cref{fig:Grafico_ej_maxflux}.

As \added[id=br]{explained} in \cite{Perez2017}, the expected maximal line flux can be estimated by the following two main factors: the linewidth set by the detector energy resolution, and the maximum allowed contribution of the line to the model, set roughly by the local error of data bins. These methods provide a reasonable \added[id=br]{and conservative} estimation for the DM flux upper limit within an energy range in which spectra are well fit by a continuum model, while remaining independent from spectral fitting of the astrophysical sources.

\begin{figure}
\centering
\includegraphics[width=0.65\hsize]{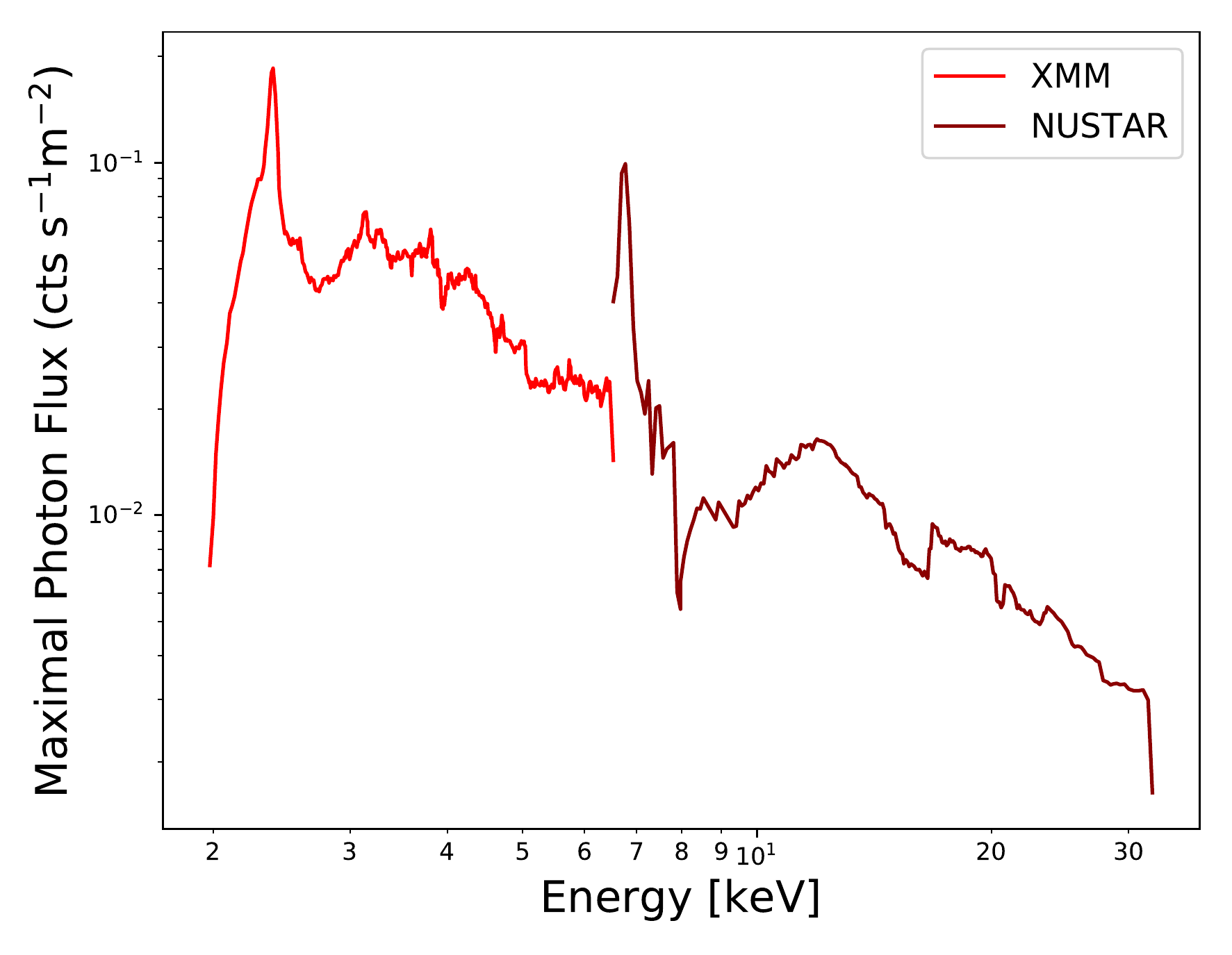}
\caption{Maximal observed flux bound for the XMM-NuSTAR observation, for $2\text{ -- }\SI{30}{\kilo\eV}$ range X-ray particle decay. We used the  $\SI{40}{\arcsec}$ circular GC region diffuse emission spectra from \cite{Mori2015}.}
\label{fig:Grafico_maxflux}
\end{figure}

Such analysis has been performed for the diffuse X-ray emission spectra observed by the NuSTAR and XMM hard X-ray surveys \cite{Mori2015}. Making use of the spectra coming from a  $\SI{40}{\arcsec}$ region around the GC, we performed this analysis for a set of energies in the range $2\text{ -- }\SI{50}{\kilo\eV}$, see \cref{fig:Grafico_maxflux}.

The maximal observed flux suffers an enhancement around $\SI{6.5}{\kilo\eV}$, due to the power law approximation of the spectra failing around neutral Fe emission lines \cite{Koyama89,Koyama96}. This will lead, in turn, to a bound relaxation in the microscopical model parameter space.

\subsubsection*{$S_{\rm DM}$ estimate}

From \cref{eq:exp_flux,eq:S_factor_definition}, the $S_{\rm DM}$ factor is obtained integrating over both the direction forming an angle $\theta$ with respect to the GC and along the line of sight, 
\begin{equation}
S_{\rm DM}=\int_{0}^{\infty}  \int_{\Omega_{\rm l.o.s.}}\rho(r(x,\theta))\,\d x \,\d\Omega ~,
\label{eq:S_factor}
\end{equation}
where  $r(x,\theta)=\sqrt{r_{\odot}^{2}+x^{2}-2r_{\odot}x \cos(\theta)}$.
We developed a systematic process in order to calculate the $S_{\rm DM}$ factor  for each density profile considered in this work. Integration for \added[id=br]{central-cored} profiles requires additional care since numerical processes yield inaccurate results for Dirac delta-like functions. The details about the calculation are discussed in \cref{appendix:SDM}. 

\begin{figure}
\centering
\includegraphics[width=0.8\hsize]{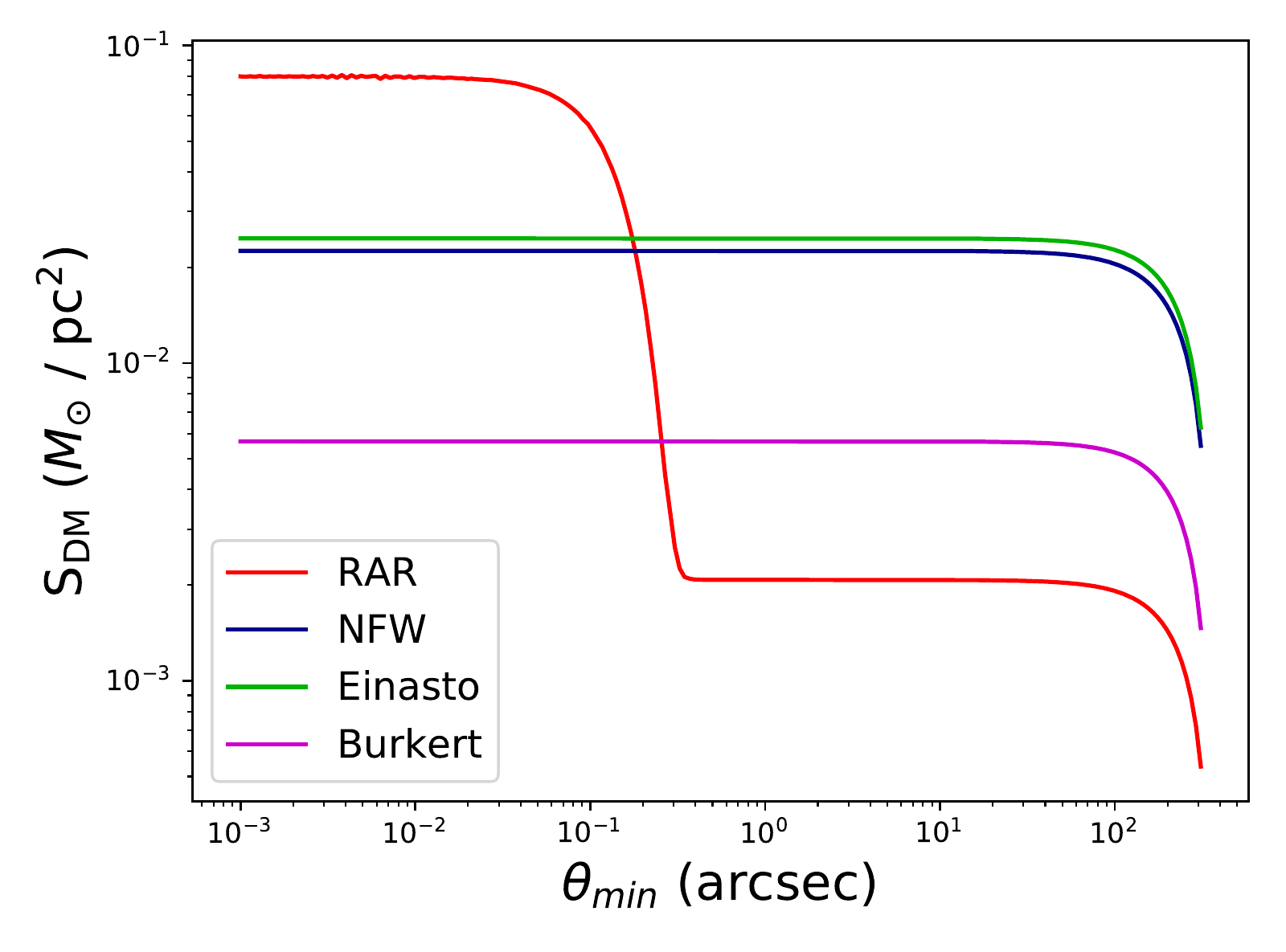}
\caption{$S_{\rm DM}$ factor for donut regions around the GC as a function of the region's minimum angle $\theta_{\rm min}$, for 4 different profiles. The maximum angle $\theta_{\rm max} = \SI{6}{\arcmin}$. NFW, EIN and BUR profiles use the parameter set specified in the text body, RAR profile uses $m_{\rm fermion}= \SI{17}{\kilo\eV}$ and parameters fitting the MW halo according to \cite{Arguelles2016}.}
\label{fig:Grafico_comp_perfiles}
\end{figure}

Our results of the $S_{\rm DM}$ factor for the four different profiles are shown in \cref{tab:S_factor_gc}. We define the GC region as a $\SI{40}{\arcsec}$ circular area around the direction of the GC. Integration is performed on the full range of the $x$ coordinate along the line of sight. 
\begin{table}
\centering
\begin{tabular}{| c | c |}
\hline
\textbf{Profile Type} & $S_{\rm DM}$ $\left[\si{\Msun\parsec^{-2}}\right]$\\
\hhline{|=|=|}
RAR                   & $\num{7.7904E-2}$   \\ \hline
NFW                   & $\num{3.6233E-4}$   \\ \hline
Einasto               & $\num{3.2055E-4}$   \\ \hline
Burkert               & $\num{6.9625E-5}$   \\ \hline
\end{tabular}
\caption{$S_{\rm DM}$ factor values for different profiles types with an integration region of $\SI{40}{\arcsec}$ circular area  around the GC. The parameters used for NFW, EIN, BUR profiles are specified in the \cref{sec:profiles}. In the case of RAR profile we consider $mc^2 = \SI{17}{\kilo\eV}$ and the parameters adopted for the Milky Way halo as in Ref.~\cite{Arguelles2016}.}
\label{tab:S_factor_gc}
\end{table}
From that results is clear that the profile choice yields important differences in this factor. As mentioned in the \cref{sec:RAR}, the RAR profiles exhibit compact cores which boost in density on several orders of magnitude at a small radius (where $r < \SI{1}{\kilo\parsec}$). Thus, we expect a significant contribution in the $S_{\rm DM}$ factor due this small section. In order to quantify such contributions, we systematically calculate $S_{\rm DM}$ factors from `donut' shaped regions of the GC integrating from different $\theta$. We show the results in \cref{fig:Grafico_comp_perfiles}. Profiles were calculated using $m_{\rm fermion} = \SI{17}{\kilo\eV}$ as a relevant example within the allowed parameter region in \cref{fig:Grafico_RAR_vs_P2017}.

The $S_{\rm DM}$ factor (thus, the expected decay photon flux) undergoes boosting once the compact core region is included. Clearly from \cref{fig:Grafico_comp_perfiles}, \added[id=br]{if one neglected this region, the factor would become much smaller with respect to the case of another profiles, thus implying less stringent limits for RAR profiles within these observation target choices}. 

\subsection{0-bounce photons}

A different observation region has been considered, this time covering a much larger portion of the observed sky, with fewer X ray luminosity (per solid angle unit). This observation follows the recent works by Perez and collaborators \cite{Perez2017} using the NuSTAR mission detectors, but aiming the analysis on the unfocused photons arriving at the detector without passing through the instrument’s focusing optics. When considering pointed observations of the GC, these photons account for the diffuse emission $\sim$~few $\SI{100}{\parsec}$ around SgrA*, however, vignetting effects due to physical blocking of the detectors by the instrument itself excludes up to the inner $\sim \SI{150}{\parsec}$, therefore reducing the astrophysical source contamination, but also removing the inner pc from the observation itself.

\subsubsection*{$S_{\rm DM}$ estimate}

For this analysis several observations are considered, roughly centered around the GC, following the procedures in \cite{Perez2017}. The total aperture from which these unfocused photons can reach the detector is about $\SI{3.5}{\degree}$ around the observation center, limited by the aperture stops attached to the focal-plane bench, and partially blocked by the NuSTAR instrument’s optics bench. These introduce both vignetting effects and physical blocking of photons arriving at the detector. Then, certain areas within the observation region are either completely blocked from detection, or the efficiency of the process is significantly diminished. Thus, to account for these effects, an efficiency factor is defined depending on the solid angle coordinates, and the S factor calculations are corrected for detector efficiency in the following form:
\begin{equation}
S_{\rm exp}=\int_{0}^{\infty}\int_{\Omega_{\rm FOV}}\epsilon(\Omega)\rho(r(x,\Omega))\d x \d\Omega ~,
\label{eq:S_factor_epsilon}
\end{equation}
with $\epsilon$ the detector efficiency factor ranging from 0 to 1.

The shape of the exposure maps for both X-ray detectors on board the NuSTAR mission are obtained in \cite{Perez2017,Wik:2014}. This sky-exposure correction factor takes into account vignetting effects and obscuration due to the instrument physically blocking photons from entering the detector from certain directions. The exposure map then excludes the inner parsecs of the GC for all observations considered here; a critical factor for dense core DM profiles \added[id=br]{as explained above}.

As an example, we calculated these factors for a Field of View of 4 degrees around the GC, for the exposure map of detector FPMA for observation \texttt{obsID 40032001002}, for three different density profiles, obtaining results as in \cref{tab:S_factor_0b}. We include coreNFW profiles in the analysis following the arguments given in \cite{Perez2017}.

\begin{table}
\centering
\begin{tabular}{| c | c |}
\hline
\textbf{Profile Type} & $S_{\rm DM}$ $\left[\si{\Msun\parsec^{-2}}\right]$\\
\hhline{|=|=|}
RAR                   & $0.3590$     \\ \hline
NFW                   & $2.0664$     \\ \hline
coreNFW               & $1.2546$     \\ \hline
\end{tabular}
\caption{$S_{\rm DM}$ factor values for different profiles types for NuSTAR mission data, using the methods described in \cite{Perez2017}. The parameters used for NFW and coreNFW profiles are specified in the \cref{sec:profiles}. In the case of RAR profile we consider $mc^2 = \SI{17}{\kilo\eV}$ and the parameters adopted for the Milky Way halo as in Ref.~\cite{Arguelles2016}.}
\label{tab:S_factor_0b}
\end{table}

Due to the exposure map suppressing the contributions form the inner parsecs of the galaxy, RAR $S_{\rm DM}$ factors are significantly suppressed and remain under the ones obtained for NFW and coreNFW. These calculations for the S factor do not include, however, possible contributions from bad pixels or ghost rays (described in \cite{Wik:2014,Mori2015}, for example). \added[id=br]{These particular features however determine} `bad data' regions and should be excluded from the observations and the S factor calculations. Both of these contributions can account for up to $\SI{70}{\percent}$ of the S factor, but are constant across profiles up to a $\SI{1.5}{\percent}$ standard deviation,\footnote{
	Tested for all profiles described in \cite{Perez2017}.
} thus remaining as an order of magnitude estimate and allowing to provide comparisons between different dark matter profiles.

\subsubsection*{Maximal Line Flux}

The joint spectra from this analysis can be seen in \cite{Perez2017} for both detectors on board the NuSTAR mission: FPMA and FPMB, as well as an in depth analysis for these signals: a recount of the astrophysical sources considered in the spectral fitting and details on the spectral reduction methods.

\begin{figure}
\centering
\includegraphics[width=0.65\hsize]{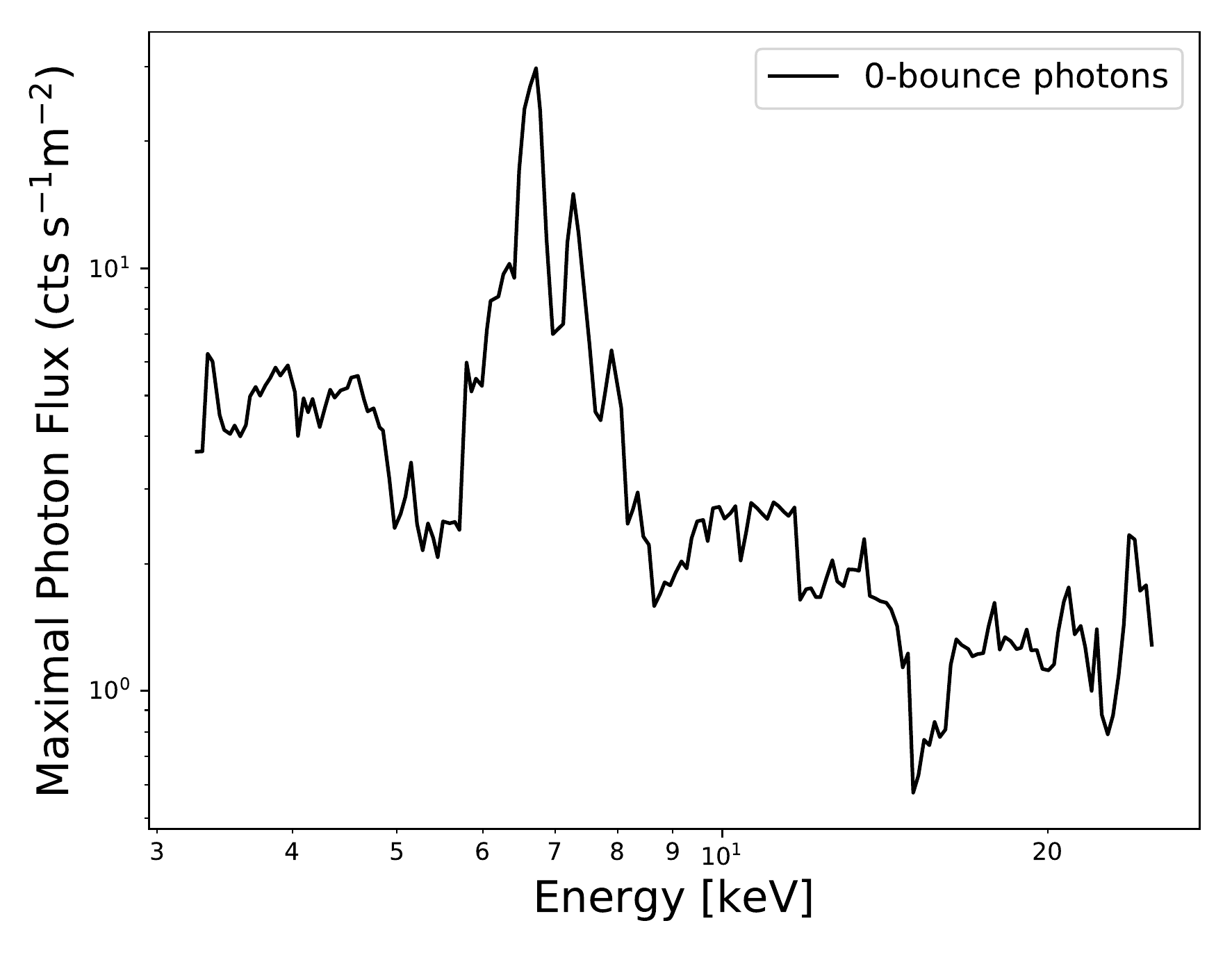}
\caption{Maximal observed flux bound for the 0-bounce photon observation, for $2\text{ -- }\SI{30}{\kilo\eV}$ range X-ray particle decay. Used the emission spectra from \cite{Perez2017}.}
\label{fig:Grafico_maxflux_P2017}
\end{figure}

We have performed the maximal expected flux analysis for the combined FPMA+FPMB spectra, and the results can be found in \cref{fig:Grafico_maxflux_P2017}. The expected flux is about a few $\si{\cts\ \second^{-1}\metre^{-2}}$, coming from a larger region of about $\SI{4}{\deg^2}$ total solid angle area.

This method of flux constraining, while it does not depend on spectral fitting models, result in overestimations when comparing with other fit dependent methods of up to a factor of a few. 

To outline quantitative differences between this method an other fit-dependent ones we compared with the results on maximal line flux obtained by \cite{Perez2017}. A $\SI{28.7}{\percent}$ mean difference excess between methods was observed within the full $3\text{ -- }\SI{25}{\kilo\eV}$ spectral range. 

\added[id=br]{It is important to stress} that an overestimation of the observed line flux leads to a relaxation in observational limits (which follows directly from \cref{eq:theta_upper_bound}) and can only result in more conservative limits for the mixing angle $\theta$. \added[id=br]{Thus, if our limits to the parameter space using RAR profiles (as exposed in \cref{fig:Grafico_RAR_vs_P2017}) would have been obtained using source-fit dependent methods in the analysis, it would lower the upper bound due to the method difference mentioned above by an average of $\sim \SI{30}{\percent}$ for this data set.}

\section{Parameter space bounds}
\label{sec:bounds} 
\subsection{Galactic Center}

Having established a limit to maximal sterile neutrino decay flux, and having calculated the theoretical expected flux, it is straightforward to obtain a $(m_{s},\theta)$ parameter space limit. Claiming that the expected flux from DM decay, eq.~\ref{eq:exp_flux}, cannot exceed the maximal decay flux coming from X-ray observations (\added[id=br]{i.e. we assume the null-detection hypothesis for this region}):
\begin{equation}
F^{\rm max}_{\rm obs}\geq F=\frac{\Gamma}{4\pi m_{s}}S_{\rm DM} ~.
\label{eq:max_flux_condition}
\end{equation}
Recalling the expression for sterile neutrino decay rate given in eq.~\ref{eq:decay_rate}, a bound on the mixing angle $\theta$ can be obtained as a function of $m_{s}$ as:

\begin{equation}
\theta^2 \leq \num{1.9465E-4}
	\left [ \frac{F^{\rm max}_{\rm obs}}{\si{\ph\ \second^{-1}\centi\metre^{-2}}}\right ]
	\left [ \frac{\si{\kilo\eV}}{m_{s}}\right ]^{4}
	\left [ \frac{\si{\Msun\parsec^{-2}}}{S_{\rm DM}}\right ]
\label{eq:theta_upper_bound}
\end{equation}

This X-ray bound becomes more stringent as more accurate constraint on maximum observed flux are achieved. Thus, non observation of DM decay lines on higher resolution equipment or tighter analytical constraints on observed flux, can only contribute to lower the bounds here obtained. The bound is also inversely proportional to $S_{\rm DM}$, so to obtain tighter constraint it is necessary to identify observational targets with boosted $S_{\rm DM}$ factors for a given DM profile. 

\begin{figure}
\centering
\includegraphics[width=0.9\hsize]{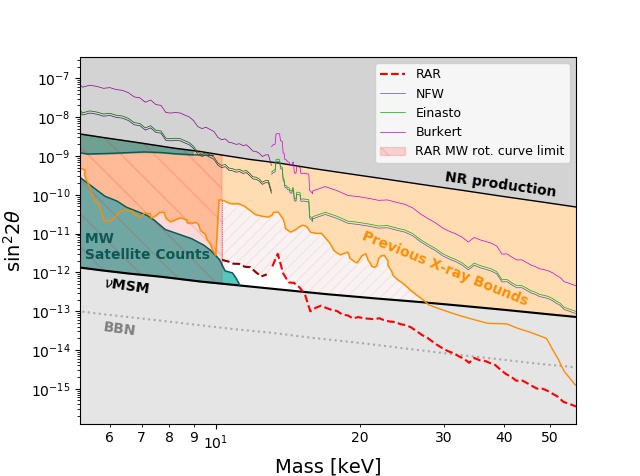}
\caption{Parameter space limits obtained for GC X-ray emission analysis. Four profile types compared: BUR, EIN, NFW and RAR. $\SI{10.4}{\kilo\eV}$ vertical shaded region correspond to RAR limits given by X ray bounds and MW rotation curve limits \cite{Arguelles2016}. Upper and lower grey shaded regions correspond to production mechanism bounds: NRP under no lepton asymmetry, $\nu$MSM for maximal model produced asymmetry and maximal BBN allowed values \cite{Boyarsky2009,DodelsonWidrow94,AsakaLaineShaposhnikov}. Shaded blue region labels MW satellite count bounds \cite{BoyarskyLowerBound,HoriuchiHumphrey}.}
\label{fig:Grafico_profile_compare}
\end{figure}

We have obtained these bounds for the mixing angle ($\theta$), and for the profiles mentioned above: NFW, Einasto, Burkert and RAR. Results can be seen in \cref{fig:Grafico_profile_compare}. Analysis has been performed for the full mass range allowed by the spectra in the case of NFW, EIN and BUR profiles. In the case of RAR type profiles, sterile neutrino masses below $\SI{10.4}{\kilo\eV}$ are disallowed, as \added[id=br]{predicted RAR (central-core) MW rotation curves start to mismatch the inner bulge data points (see for example the clear overshooting for the case of $m_{s}=\SI{0.6}{\kilo\eV}$ in \cref{fig:RAR-MW})}.

\begin{figure}
\centering
\includegraphics[width=0.7\hsize]{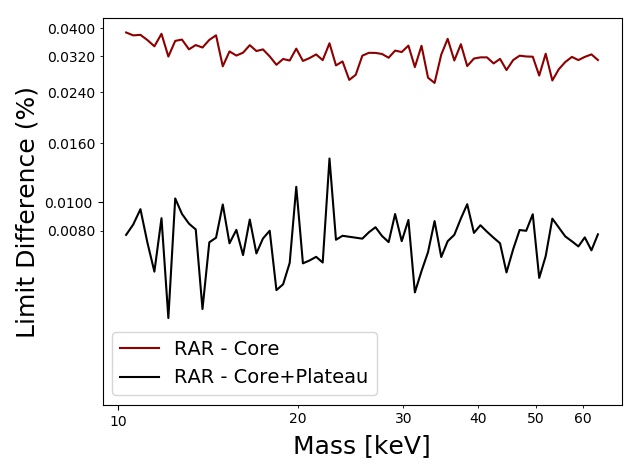}
\caption{X-ray limit contribution for different sections of RAR profiles, shown as a relative difference between reduced RAR (core and core+plateau) and full RAR profiles.}
\label{fig:Grafico_comp_RAR}
\end{figure}

As expected, the $S_{\rm DM}$ factor enhancement for \added[id=br]{the novel core-halo RAR type} of profiles results in lower limits in parameter space. As seen before, this enhancement results from the inclusion of the central regions of the GC in the observation, which increases $S_{\rm DM}$ factors and brings them over the ones arising from the other three profiles.

We also analyzed the main contributions to the $S_{\rm DM}$ factor for the RAR profiles. A RAR profile can be split into three distinct features: a compact high density quantum core at $r \lesssim \SI{1}{\parsec}$, a baryonic dominated plateau at $r \approx 1\text{ -- }\SI{E4}{\parsec}$ and an outermost halo at $r > \SI{E4}{\parsec}$. We plot the contributions to the X-ray sterile decay limit from profiles comprised only of the core component, core+plateau and a full RAR profile on \cref{fig:Grafico_comp_RAR}, calculated as $|(\theta_{c/c+p}-\theta)/\theta|$, with $\theta_{c/c+p}$ the limit contributions from core and core+plateau only profiles. We find no significant difference between the limits, thus a $S_{\rm DM}$ factor from GC observation can be approximated by the one obtained from only considering the compact core for RAR profile calculations.


\subsection{0-bounce photons}

The \added[id=br]{parameter-space} bound analysis for the 0-bounce photons spectrum is very similar to the one previously mentioned. A few differences reside in the calculation for the $S_{\rm DM}$ factors.

Namely, the exposure map corrections mentioned in previous sections, in addition to averaging over different observations. As the spectrum has been averaged over six observations, and co-added for FPMA and FPMB, each with different exposure maps, the expected flux must be obtained via a weighted average of $S_{\rm DM}$ factors for each one of the observations.

The $S_{\rm DM}$ factor has been calculated as:
\begin{equation}
S_{\rm avg}=\frac{\sum_{i} \Delta t \Delta \Omega S_{i}}{\sum_{i}\Delta t \Delta \Omega} ~.
\label{eq:S_factor_average}
\end{equation}
With $\Delta t$ the exposure time and $\Delta \Omega$ the effective detector area for each observation. The specific values of these parameters and \added[id=br]{further observation details} can be found in \cite{Perez2017}.

\begin{figure}
\centering
\includegraphics[width=0.8\hsize]{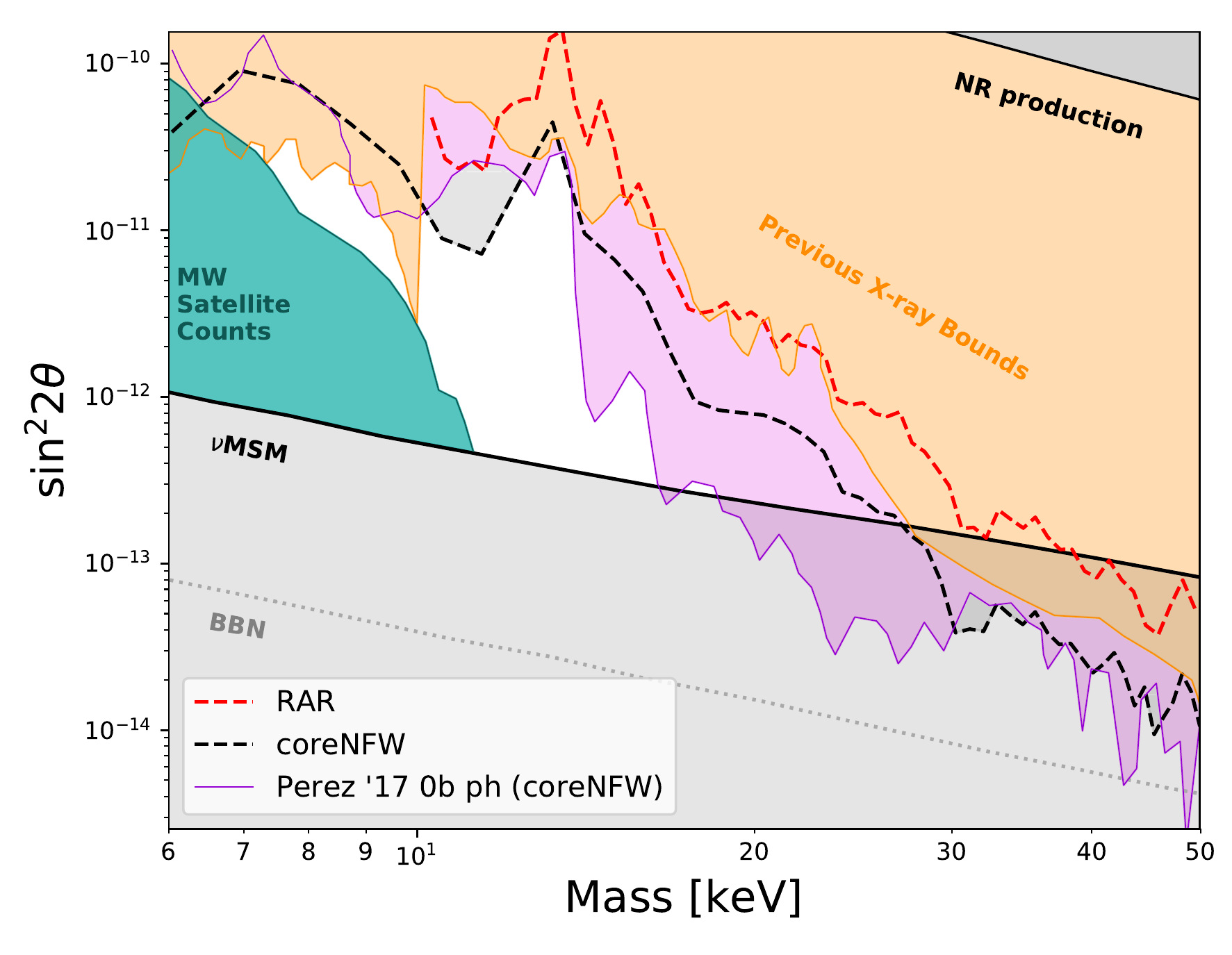}
\caption{Parameter space limits obtained for 0-bounce photon analysis. Two profile types compared: coreNFW and RAR. Violet line refers to analysis by \cite{Perez2017}, corresponding to a coreNFW profile.}
\label{fig:Grafico_profile_compare_P2017}
\end{figure}

Once these averages have been taken, the procedure for obtaining a bound for $\theta$ are similar to the one taken for the GC. We performed the analysis for profiles NFW and RAR (with parameters previously mentioned), obtaining the results in \cref{fig:Grafico_profile_compare_P2017}.

The limits for the RAR profile are significantly relaxed for this observation region. This is to be expected, as inner compact regions for these profiles are excluded from the exposure map, therefore not contributing to the $S_{\rm DM}$ factor and thus relaxing the expected bounds. 


\subsection{\texorpdfstring{$\nu$}{nu}MSM parameter bounds from other sources}
\label{sec:comparison}

We showed \added[id=br]{sterile neutrino free-parameter} constraints obtained for different regions around the GC, and compared different profiles for each region. All the constraints shown here are upper bounds on the mixing angle as a function of the particle mass, but several other limits exist that limit the parameter space for the \added[id=br]{particle model under consideration}. Phase space density constraints place a lower bound on mass at around $1\text{ -- }\SI{2}{\kilo\eV}$, and further X-ray explorations rule out particle masses above $\SI{70}{\kilo\eV}$ (see \cite{Perez2017} and references therein). Also, production mechanisms place lower bounds on the mixing angle, leaving a small window of allowed parameter combinations. Recent works \cite{NoMSP2017} provide more stringent lower mass bounds for WDM models based on MW satellite counts, where particle masses lower than $\SI{4}{\kilo\eV}$ are strictly ruled out, and suggesting possible future limits for the DM particles around $\SI{8}{\kilo\eV}$.

Lower bounds on the production angle, labeled in the various parameter space plots as $\nu$MSM, BBN and NRP arise from sterile neutrino production mechanisms in the early universe by mixing with the standard sector. These mechanisms are heavily dependent of the values for lepton asymmetry in this early stage: Non-Resonant Production (NRP) \cite{DodelsonWidrow94,AsakaLaineShaposhnikov,2017JCAP...01..025A} mechanisms are the only ones to take place in the absence of asymmetry, and assuming overclousure  of the universe a relation between $m_{s}$ and $\theta$ can be plotted in the case of this symmetric universe. The presence of lepton asymmetry adds another available production mechanism: Resonant Production (RP) \cite{LaineShaposhnikov,ShiFuller,AsakaLaineShaposhnikov}. Part of the DM abundance can be generated in this way, thus relaxing the limits on the mixing angle and allowing lower values. In the context of $\nu$MSM lepton asymmetry can be produced via decays of heavier particles, up to the value that outlines the line labeled as $\nu$MSM in the parameter space plots. If we remain agnostic to the origins of this asymmetry, limits can be further lowered until these values come into conflict with nucleosynthesis predictions, a limit labeled as BBN. Lower bounds arising from production mechanisms are further discussed in \cref{sec:interactingDM}, in the context of a \added[id=br]{minimally-extended $\nu$MSM theory}, \added[id=br]{allowing for a new production mechanism.} 

\subsection{Comparison with recent works}
\label{sec:comparison2}

Perez et al. \cite{Perez2017} conducted similar searches using 0-bounce photons, and lowered these upper mixing angle bounds further than previous works. In the past sections, we showed that for such analysis and observation region, no stronger limits could be obtained from RAR profiles. However, for observations that include the inner parsecs of the GC, we found that RAR profiles provide stronger limits than NFW counterparts. We then compare the limits obtained for GC observations with the limits obtained by \cite{Perez2017} in \cref{fig:Grafico_RAR_vs_P2017}.

RAR profiles for these observations then result in stronger upper bounds for the mixing angle. Under this analysis, the parameter space window for the full $\nu$MSM model is further constrained for RAR profiles. We recall (see discussion in \cref{sec:signal}) that fit-dependent maximal observed flux, will certainly set more stringent limits which may completely rule out the $\nu$MSM model, under RAR model assumption for galactic DM. 

However, some minimal modifications to the $\nu$MSM which include for interactions in the dark sector, may likely left open the relevance of sterile neutrinos to play the role of the DM in the Universe. Indeed, one should bear in mind that the lowest $\nu$MSM parameter bounds are subject to the sterile neutrino production mechanism, which in turn is sensitive to the adopted rate between lepton-to-baryon asymmetry in the Universe (see \cite{BoyarskyLowerBound} for details). For example, if one is agnostic to the generation of lepton asymmetry within $\nu$MSM, then only remains the lower BBN limit, which directly constrains the maximum amount of lepton asymmetry allowed in the Early Universe. Interestingly, novel generation mechanisms for our fermionic candidates beyond $\nu$MSM, will certainly relax such lower bounds, opening the possibility for new sterile neutrino physics as the cosmological DM. Further inspection of these matters are discussed in next section.

\section{Parameter space relaxation \& dark sector interactions}
\label{sec:interactingDM}

The above described constraints refer exclusively to the case where the \added[id=br]{DM fermions} are identified with the right handed neutrinos of the $\nu$MSM model~\cite{nuMSM}.  The constraints may be relaxed if (\added[id=br]{minimal}) generalisations of the model are considered, assuming self-interactions among the \added[id=br]{particles}, induced, for instance, by the
exchange of massive vector particles in a dark sector of the model. \added[id=br]{Such an extension was proposed} in \cite{amrr} in an attempt to explain the observed \added[id=br]{rotation curves (from center to periphery)}, as well as alleviate discrepancies between observations at galactic (small cosmological) scales and predictions based on numerical simulations based on the $\Lambda$CDM model (``small-scale Cosmology crisis'')~\cite{2017ARA&A..55..343B}.
Interactions with pseudoscalars (axion-like excitations) in either visible~\cite{pilaftsis} or dark~\cite{dk,chuda} sectors of the pertinent theory have also been considered. In particular, in \cite{pilaftsis}, Yukawa type interactions of right-handed neutrinos with axion pseudoscalars have been proposed as a novel mechanism for generating a Majorana mass for the right handed neutrinos beyond seesaw~\cite{1977PhLB...67..421M,1980PhRvL..44..912M,1980PhRvD..22.2227S,1981NuPhB.181..287L,1979GellMann,1979Yanagida}, and in this sense they can also be included in the model of \cite{amrr} in addition to the vector interactions, thus significantly affecting the pertinent constraints.

The presence of such extra ingredients, entail also the important hint that dark matter may consist not only of a dominant component, but of several species, playing a different r\^ole at various scales. This relaxes significantly constraints on mixing parameters between \added[id=br]{the DM candidates} and Standard Model matter arising from the requirement of avoiding overclosure of the Universe.
It is the purpose of this section to discuss briefly several such scenarios and how they modify/relax the constraints pertaining to the $\nu$MSM model of \cref{fig:Grafico_RAR_vs_P2017}.

\subsection{Vector Interactions \added[id=br]{among sterile neutrinos}}
\label{vector}

We commence our discussion by considering the \added[id=br]{self-interacting DM} model of \cite{amrr}. The relevant Lagrangian is given by :
\begin{equation}
{\mathcal L}={\mathcal L}_{GR}+{\mathcal L}_{N_{R\,1}}+{\mathcal L}_V+{\mathcal L}_{I}\,
\label{eq:Ltotal}
\end{equation}
where
\begin{align}
{\mathcal L}_{GR} &= -\frac{R}{16\pi G}, \nonumber \\
{\mathcal L}_{N_{R\,1}} &= i\,\overline{N}_{R\, 1}\gamma^{\mu}\, \nabla_\mu\,N_{R\,1}-\frac{1}{2}m\,\overline{N^c}_{R\, 1}N_{R\,1}, \nonumber \\
{\mathcal L}_V &= -\frac{1}{4}V_{\mu\nu}V^{\mu\nu}+\frac{1}{2}m_V^2V_{\mu}V^{\mu},\nonumber \\
\label{int}
{\mathcal L}_{I} &= -g_V V_\mu J_V^\mu=-g_V V_\mu \overline{N}_{R\, 1}\gamma^{\mu}N_{R\,1}\, ,
\end{align}
with $R$ the Ricci scalar for the metric background, which, for the purposes of \cite{amrr} which was the study of galactic profiles, it was assumed static  and spherically symmetric: $g_{\mu \nu}={\rm diag}(e^{\nu},-e^{\lambda},-r^2,-r^2\sin^2\varphi)$; with $e^{\nu}$ and $e^{\lambda}$ functions only of the radial coordinate $r$, and $\varphi$ denotes the polar angle. The quantity $\nabla_\mu=\partial_\mu\, -\, \frac{i}{8}\, \omega_\mu^ {ab}[\gamma_a, \gamma_b] $ is the gravitational covariant derivative acting on a spinor, with $\omega_{\mu }^{ab}$ the spin connection; $m$ is the Majorana mass of the sterile neutrino, whose microscopic origin was left unspecified in \cite{amrr}.
The right-handed sterile neutrinos are denoted by $N_{R\,1}$ \added[id=br]{and the superscript $`c$ over a spinor field denotes the charge conjugate, satisfying the Majorana four-spinor condition, ${\mathcal N}^c={\mathcal N}$ (see \cite{amrr} for further model details and properties).} 

For simplicity we assume  minimal-coupling of the vector field with the sterile neutrino current $J^\mu_V$ in the
interaction term ${\mathcal L}_{I}$ (\ref{int}).  This current is conserved if decays of sterile neutrinos are ignored, as done in \cite{amrr}.


Galactic phenomenology \added[id=br]{accounting for self-interactions as in~\cite{amrr}, implies modified Ruffini-Arg\"uelles-Rueda (RAR) profiles (see~\cite{RAR1,Arguelles2016,siutsou,ars} for standard ones), with \textit{compact core-diluted halo} profiles, including  for sufficiently dense cores in order to provide an alternative to massive BHs in agreement with overall rotation curve observations. Such an agreement is acquired for a family of RAR solutions with a corresponding minimum mass of the right handed neutrino about $\SI{47}{\kilo\eV}$ (with its maximum allowed mass of about $\SI{350}{\kilo\eV}$, to avoid gravitational collapse)}. This particle mass based on considerations of the structure of the Milky Way and of other galaxy types, does not fall within the allowed narrow mass range found in this work from an indirect detection analysis within the $\nu$MSM standard scenario. However, when one relaxes the condition that the central quantum core be an alternative to the massive BH scenario, the RAR profiles (either with or without DM self-interactions) can still fit the overall rotation curve in galaxies allowing for lower particle masses down to $\sim 10$~keV\footnote{We recall that in \cite{Arguelles2016} it was found that a particle mass in the range $10\text{ -- }\SI{48}{\kilo\eV}$ is also allowed by the Milky Way rotation curves data. However, in that case the DM quantum core cannot explain the dynamics near the Galactic center so it does not provide an alternative to the BH scenario for SgrA*. In that particle mass range one could consider the DM distribution obtained in a BH$+$RAR-halo model which could differ near the Galactic center with respect to the one presented in \cite{Arguelles2016}. This calculation is out of the scope of the present work and will be presented elsewhere.}. Importantly, all these results have been obtained assuming a negligible mixing of the sterile neutrino with the SM sector. The case where such a mixing is turned on is discussed in this work through an indirect detection analysis but without allowing for self-interactions among the DM particles. Possible effects and consequences for the full case including for both kind of interactions (though not solved here) are outlined at the end of this section.

In addition, if one insists that the vector-field-induced self-interactions in the sterile neutrino sector provide solutions to the small-scale cosmology crisis~\cite{2017IJMPD..2630007M}, then a strong cross section relative to that of the conventional weak interactions in the SM sector, is required~\cite{amrr}. Indeed, to resolve such tensions between predictions of $\Lambda$CDM-based numerical simulations and observations, the self-interacting DM (SIDM) cross section has to be in the range~\cite{harvey}
\begin{equation}
0.1 \, \le \frac{\sigma_{\rm SIDM}/m}{{\rm cm}^2\, g^{-1}} \, \le 0.47~,
\label{newconstr}
\end{equation}
according to recent measurements employing novel observables of colliding galaxies. The vector interactions (\ref{int}) in our case, lead to a cross section
\begin{equation}
\sigma^{\rm tot}_{\rm core}\approx\frac{(g_V/m_V)^4}{4^3\pi}29m^2 \qquad (p^2/m^2\ll 1)\, .
\label{eq:20}
\end{equation}
which, on account of \cref{newconstr}, imply for the strength of the vector interaction of the sterile neutrino sector relative to the Fermi coupling $G_{\rm F}$ of the SM weak interactions~\cite{amrr}
\begin{equation}
\overline{C}_V \equiv \left(\frac{g_V}{m_V}\right)^2 G_{\rm F}^{-1} \in (\num{2.6E8},\num{7E8}),
\end{equation}
for ino masses in the range $mc^2 = 47\text{ -- }\SI{350}{\kilo\eV}$, implying that the mass of the massive-vector meson would be constrained to values $m_V\lesssim \SI{3E4}{\kilo\eV}$, in order to satisfy $g_V\lesssim 1$ as requested by the self-consistency of the perturbation scheme we have applied to compute the cross section by \cref{eq:20}.\footnote{
	We recall that for the case of particle masses below $\SI{47}{\kilo\eV}$, as the central core within the RAR model does not provide an alternative to the central BH (see~\cite{Arguelles2016,amrr}), the above values are somewhat modified.
} 

The presence of the vector-sterile-neutrino interaction term ${\mathcal L}_I$ plays an important r\^ole in the relaxation of the constraints of \cref{fig:Grafico_RAR_vs_P2017}, since it implies an additional production channel for the DM sterile neutrino $N_{R\, 1}$ through the decays of the massive vector field $V_\mu$ in the early universe. So, sufficient production of DM may be guaranteed even if one ignores any coupling of sterile neutrinos with the SM sector, by assuming negligible Yukawa couplings $F_{\alpha\, 1}$ (\added[id=br]{see \cref{yuk_s2}}).

Indeed, as discussed in Appendix, the rate of decay (width $\Gamma$) of the vector Boson into a pair of identical Majorana particles (whose mass  $ \simeq \mathcal{O}(50)~\si{\kilo\eV}$
is viewed as negligible when compared to that of the
boson $V_\mu$, $m_V \simeq \SI{1E4}{\kilo\eV}$, according to the phenomenological analysis of \cite{amrr} in order to reproduce the observed galactic structure) is given approximately at tree level by
\begin{equation}\label{vnuwidth}
\Gamma_1   \simeq \frac{g_V^2} {48\, \pi}\, m_V~.
\end{equation}
Quantum corrections may affect this result, but will not be the focus of our brief discussion in this work. In models with more than one generation of right handed neutrinos coupled to the vector field there are extra contributions to the total width, which amount to a multiplication of the result in \cref{vnuwidth} by the number of right-handed neutrino flavours $N_f$ (usually $N_f=3$, like in the case of $\nu MSM$~\cite{nuMSM}).

The freeze-out temperature of the reaction is estimated by equating $\Gamma_1$ in \cref{vnuwidth} with the Hubble parameter $H$ of the Universe, $\Gamma_1 = H$.
Assuming standard cosmology, in which there is radiation dominance in the Early Universe, the Hubble parameter is expressed in terms the temperature $T$ as~\cite{kolb}
\begin{equation}\label{hubble}
H=1.66 \, T^2 \mathcal{N}^{1/2} M_{\rm Pl}^{-1}~,
\end{equation}
where $\mathcal{N}$ is the effective number of degrees of freedom of all elementary particles and $M_{pl}$ is the reduced Planck mass. For a minimal extension of the SM, with only right-handed neutrinos and the background $B_0$, we may estimate $\mathcal{N} = \mathcal{O}(100)$ at temperatures higher then the electroweak transition.
Equating \cref{vnuwidth} with \cref{hubble} we obtain for the pertinent freeze-out temperature, $T_{D}$,
\begin{equation}\label{freeze}
T_{D1}\simeq 6.3\cdot 10^{-2} \frac{|g_V|}{\mathcal{N}^{1/4}}\sqrt{m_V \, M_{\rm Pl}}
\end{equation}
As discussed above, the requirement of alleviating the small-scale cosmology crisis via these vector-sterile-neutrino interactions requires~\cite{amrr} $m_V \simeq \SI{10.4}{\kilo\eV}$, with $g_V = {\mathcal O} (1)$; we then obtain from \cref{freeze} that $T_D = {\mathcal O}(10^8)~\si{\giga\eV}$, which yields the ball park of temperatures in which the sterile neutrino DM abundance is created in our interacting DM model.

The calculation of the sterile-neutrino thermal abundance at the freeze-out can be done as usual by the solution of the pertinent system of Boltzmann equations, or better out of equilibrium thermal field theory techniques (e.g. Kadanoff-Baym equations).
In general, one may end up with overproduction of sterile neutrino dark matter that would lead to overclosure of the Universe, unless the would-be freeze-out temperature of the vector mesons
lies above the reheating (or even preheating) temperature of the Universe. The latter is not known but it might be constrained by some CMB observations, with a lower limit lying
in the range $\sim 20\text{ -- }900$~\cite{preheat1,preheat2}. We observe that in our simplified model the freeze-out temperature in \cref{freeze} is much higher than such lower limits of reheating temperature, and hence overproduction of warm sterile neutrino DM, through the decays of the vector boson, might be achieved. Other ways of avoiding overproduction of DM is via the dilution of the relic right-handed neutrino density by release of entropy through. e.g. decays of the heavier right-handed neutrinos (in models with more than one generation of sterile neutrinos) after their freezeout~\cite{rhnsu2}.


The addition to \cref{eq:Ltotal} of a Yukawa Higgs-portal term as explicited above in \cref{yuk_s2}, changes the situation drastically. Indeed, as we already discussed, upon considering such a coupling, one obtains the stringent X-ray and BBN constraints of the mixing angle and mass of $N_{R\, 1}$ depicted in \cref{fig:Grafico_RAR_vs_P2017}, given that \cref{yuk_s2} implies decays of the heavy neutrinos $N_I \rightarrow \nu H$, where $H$ denotes the Higgs excitation field, defined via: $\phi = \langle \phi \rangle + H $. In such a case $J^\mu_V$ is \emph{not} conserved in time. However, in the context of $\nu$MSM, the lightest of the heavy neutrinos decay time is longer than the age of the universe~\cite{nuMSM}, hence the latter can be considered as stable for all practical purposes, thus playing the r\^ole of a DM component.


The thermal history of the Universe in the combined model where both the interaction term (\ref{int}) and the mixing (\ref{yuk_s2}) is more complicated and we shall not present it here. However,
the Dirac Yukawa coupling  of the mixing term given by \cref{yuk_s2} for a $\si{\kilo\eV}$ sterile neutrino, of interest here, is sufficiently weak (as required by the seesaw scenarios~\cite{1977PhLB...67..421M,1980PhRvL..44..912M,1980PhRvD..22.2227S,1981NuPhB.181..287L,1979GellMann,1979Yanagida} of generating a light active neutrino mass in the SM sector) so it cannot bring the sterile neutrinos into thermal equilibrium above electroweak-scale temperatures. So, in the presence of our vector interactions with a freezeout of order (\ref{freeze}) they will not play a dominant r\^ole in the sterile neutrino abundance.

\section{Conclusions}
\label{sec:concls}

We have investigated the X-ray signal expected from the GC due sterile neutrino decays in the energy range $E_{\gamma} = 2\text{ -- }\SI{50}{\kilo\eV}$. The intensity of the DM decay flux expected from an individual halo depends mainly on the DM density distribution inside it. To discuss in detail the theoretical uncertainties in the calculation, we have estimated the signal assuming different DM densities profiles. Concretely, in \cref{sec:decay} we have described the signal with the decay width defined assuming the $\nu$MSM model and presented different parametrizations such as NFW, Burkert, Einasto. In addition, the RAR profile was considered in the analysis which has the distinct feature of depending on the particle mass. Our results for the $S_{\rm DM}$ factor considering the different profiles are presented in \cref{sec:signal}.  We conclude that the profile choice yields important differences in this factor as expected. Since the RAR profiles exhibit compact cores which boost in density at small radius we obtain the maximum value for the $S_{\rm DM}$ factor with respect to the other profiles when the compact core region is included. However, we obtain lower values assuming the RAR profile when such regions are not considered in the calculation. The dependence of the $S_{\rm DM}$ factor with the integration of the minimum value of the angle $\theta_{\rm min}$ forming with the GC direction, is shown in the \cref{fig:Grafico_comp_perfiles}.

In order to calculate new constraints on the parameter space for sterile neutrino DM models, we compare our analytical results with the X-ray flux observations from the GC and 0-bounce photons taken by the NuSTAR and XMM missions. In \cref{sec:signal} we describe our signal analysis for the GC region (\added[id=br]{at few pc scales from SgrA*}) and also the use of NuSTAR 0-bounce photons (\added[id=br]{corresponding to $\sim \SI{E2}{\parsec}$ off the Galaxy center}). We use the NFW and the RAR profile as the representative of our final results about the limits on the mixing angle for sterile neutrino DM in \cref{sec:bounds}. Assuming that no signal is observed (\added[id=br]{i.e. we assume a null-detection hypothesis}), in \cref{eq:theta_upper_bound} we discuss the dependencies of the bound respect to $m_s$, $S_{\rm DM}$ and $F_{obs}$. For the GC region, the lower limit in the mixing angle is obtained when the RAR profile is considered since the bound is inversely proportional to $S_{DM}$. \added[id=br]{Such novel lower bound imply the most stringent constraints up to date on the parameter space, with an allowed particle mass window narrowed down $m_s\sim 10\text{ -- }\SI{15}{\kilo\eV}$ within the $\nu$MSM sterile neutrino model (see \cref{fig:Grafico_RAR_vs_P2017}), possessing a radiative decay channel into X rays}. The upper one corresponds to the BUR profile (see \cref{fig:Grafico_profile_compare}). In the case of use the 0-bounce photons, the RAR profile results in stronger upper bound since the inner compact regions are excluded. 
  
In addition, we further discuss on the possible effects in the sterile neutrino parameter-space bounds due to a self-interacting nature of the dark matter candidates. Self interacting models have been reviewed in \cref{sec:interactingDM}, that arise from a generalisation of $\nu$MSM by introducing an exchange of massive vector particles in the dark sector. We briefly consider the implications of such types of models in sterile neutrino physics as cosmological DM, which may circumvent the requirement for universe overclousure, and produce the required DM abundance via dark sector interactions without imposing strong limits to the standard model-dark matter mixing angle $\theta$ as those seen in \cref{fig:Grafico_RAR_vs_P2017}.

\acknowledgments
The work of C.R.A is supported by the National Research Council of Science and Technology (CONICET-Argentina).
C.R.A and A.M. further acknowledges the hospitality of the ICRANet Headquarters, where finalization of the present work was taking part.
The work of N.E.M. is partially supported by the
U.K. Science and Technology Facilities Council (STFC) via the Grant ST/L000258/1.
N.E.M. also acknowledges currently the hospitality of IFIC Valencia through a Scientific Associateship (\emph{Doctor  Vinculado}).
A.K. is supported by the Erasmus Mundus Joint Doctorate Program by Grants Number 2014--0707 from the agency EACEA of the European Commission.
The research of A.M. is supported by Fundação para  a  Ciência  e  a  Tecnologia  (FCT)  through  national funds  (UID/FIS/04434/2013)  and  by  FEDER  through COMPETE2020 (POCI-01-0145-FEDER-007672). 

\clearpage

\appendix
\section{S factor}
\label{appendix:SDM}

An algorithm was developed to perform the $S_{\rm DM}$ factor integral, comprising of a solid angle integral and an integral along the line of sight, as seen in \cref{eq:S_factor}.

Each integral is performed as a Riemann sum: an integral is approximated as the sum of the function values on a grid, times the spacing between elements of such grid, as in:
\begin{align}
&\int_{a}^{b} f(x) \d x \approx \sum_{i=1}^{n}f(x_{i})\Delta x_{i} \\
&\text{where}\ \ x_{i}\in [a,b] \ ,\ \Delta x_{i}=\frac{x_{i+1}-x_{i-1}}{2} \nonumber
\end{align}
And the process is trivially extended for double and triple integrals. On the limit $n\rightarrow\infty$ both expressions are equivalent if $x_{i}$ are evenly spaced between a and b.

This kind of approximations yield greater errors in areas where $f$ changes rapidly and $X_{i}$ evaluation points are scarce. Thus, it is critical for the accuracy of these algorithms to make a good choice of evaluation points $X_{i}$. We will start by analyzing the solid angle integral, and how it is possible to optimize the evaluation points for the Riemann sum.

First, spherical symmetry of the halo density profile can be used to evaluate the dependence on the azimuthal angle $\Phi$:
\begin{equation}
\int_{\theta=0}^{\theta_{\rm max}}\int_{\Phi=0}^{2\pi}\bar{S}(\theta,\Phi)\sin{\theta}\d\theta \d\Phi=2\pi\int_{\theta=0}^{\theta_{\rm max}}\bar{S}(\theta) ~.
\end{equation}
Then only remains to choose a suitable choice of evaluation points for $\theta$. For circular shaped regions we chose logarithmically spaced points between $\theta \approx \SI{E-6}{\arcsec}$ and $\theta_{\rm max}$ ($\SI{40}{\arcsec}$).\footnote{The lower angle corresponds to the shortest relevant radius for RAR profiles.
} This allows us to have a greater definition around the profile inner regions, where density is expected to change rapidly. Logarithmic spacing was also used for `donut' shaped regions mentioned in \cref{sec:SA_GC}.

A more complex analysis is required for the integral along the line of sight:
\begin{equation}
\bar{S}(\theta,\Phi)=\int_{x=0}^{x=\infty}\rho(r(x,\theta,\Phi))\d x ~.
\end{equation}
Here we find the same challenge in evaluation pints: it is necessary to have tightly packed points around the $x$ values closer to the GC. But other numerical problems arise in the definition of r: 
\begin{equation}
r=\sqrt{r_{\rm GC}^{2}+x^{2}-2xr_{\rm GC}\cos{\theta}} ~.
\label{eq:r_one}
\end{equation}
Where $r_{\rm GC}$ is the distance between the Sum and the GC ($\approx \SI{8E3}{\parsec}$). If parameters are such that we can access the inner regions of the halo profile ($\approx \SI{E-7}{\parsec}$), then this result is to be acquired from the subtraction of two similar quantities up to $\SI{E-14}{\kilo\parsec^2}$: $(r_{\rm GC}^2+x^2)$ and $2xr_{\rm GC}\cos{\theta}$. This would have resulted in floating point precision errors for ordinary data storage types. Then, it was necessary to find an expression for $r(x,\theta)$ that remained accurate on such scales.

\begin{figure}
\centering
\includegraphics[width=0.65\hsize]{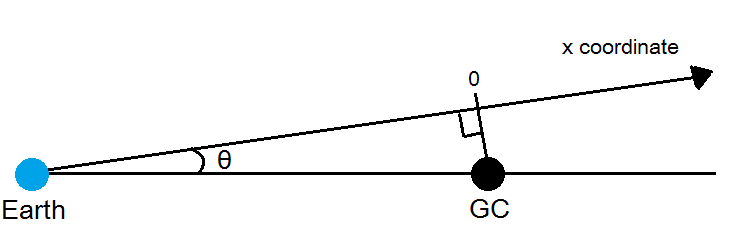}
\caption{Coordinate system election schematic. $x-\theta$ plane slice. X coordinate zero is set at the axis' closest point to the GC.}
\label{fig:Esquema_x_coord}
\end{figure}

So, we first redefine the zero of the x coordinate, so it is measured from the closest point to the GC, as shown in \cref{fig:Esquema_x_coord}.
Then, for small theta the expression in \cref{eq:r_one} can be approximated as:
\begin{align}
&r^{2}=r_{\rm GC}^{2} \left ( -2\epsilon(\theta) + \left( \frac{x}{r_{\rm GC}}\right )^2 - 4\frac{x}{r_{\rm GC}}\epsilon(\theta)\right )\nonumber\\
&\text{where}\  \epsilon(\theta)=\cos{\theta}-1=-\frac{\theta^2}{2}+\frac{\theta^4}{24}-...
\label{eq:r_2}
\end{align}
Where, for small $\theta$ and positive x it involves sums of positive expressions only.

Using this new definition of the x coordinate origin, we can solve the sampling problem using logarithmic spaced evaluation points around x=0. This kind of spacing was used on the intervals $X=[-r_{\rm GC},-10^{-7}]$ and $X=[10^{-7}, 2 r_{\rm halo}]$, with $r_{\rm halo}$ the MW DM halo radius ($\approx \SI{55}{\kilo\parsec}$). Then, Riemann sums were executed to evaluate the integral on these points using \cref{eq:r_2}.

\section{A sterile neutrino production mechanism: V-boson decay}

The required Decay rate of a $V$-boson decaying into two sterile neutrinos, in view of the interaction term (\ref{int}) is given by the standard formula:
\begin{equation}\label{width}
\Gamma = \frac{{\cal S}}{2E}\left[\int \prod_{i=1}^{m}\frac{\d^3p'_i}{2E_i' (2\pi)^3}\right]|\overline{{\cal M}}|^2 (2\pi)^4 \delta^{(4)}(\sum_{i=1}^m p_i' - p) ,
\end{equation}
where $p$ ($p_i^\prime, \, i=1, \dots m$) denote the four- momenta of the decaying particle (decay products),
${\cal S}$ is a statistical factor that equals $1/(n!)$ for each group of $n$ identical particles in the decay products, and $|\overline{{\cal M}}|^2$ is the square of the matrix element between initial and final states, averaged over initial-spin states and summed over final-spin ones.

For the evaluation of the amplitude ${\cal M}$ we use the Feynman rule for the vertex $V_\mu \, \overline{\nu_{1}}\,  \nu_{1} $
\begin{equation}\label{rules}
- i g_V\, \gamma^\mu \, \Big( \frac{1 + \gamma_5}{2}  \Big) ~.
\end{equation}
To evaluate the decay width (\ref{width}) we use the following Casimir identities (in the following expressions, $u_a(p)$ ($v_b(p)$) denote polarization spinors (antispinors) appearing
in the solution of the free Dirac (or Majorana) equations):
\begin{align} \label{rules2}
&\sum_{i_A,i_B=\uparrow,\downarrow} \left({\overline u}^{(i_A)}(p_A)\Gamma_I u^{(i_B)}(p_B)\right)^\dagger \left({\overline u}^{(i_A)}(p_A)\Gamma_{II} u^{(i_B)}(p_B)\right)
\nonumber \\
&= {\rm Tr}\left[{\overline \Gamma}_I \left(\gamma^\mu {p_A}_\mu  + m_A\right)\Gamma_{II} \left(\gamma^\mu {p_B}_\mu + m_B\right)\right], \nonumber \\
& \sum_{i_A,i_B=\uparrow,\downarrow} \left({\overline v}^{(i_A)}(p_A)\Gamma_I u^{(i_B)}(p_B)\right)^\dagger \left({\overline v}^{(i_A)}(p_A)\Gamma_{II} u^{(i_B)}(p_B)\right) \nonumber \\
&= {\rm Tr}\left[{\overline \Gamma}_I \left(\gamma^\mu {p_A}_\mu  - m_A\right)\Gamma_{II} \left(\gamma^\mu {p_B}_\mu + m_B\right)\right] , \nonumber \\
& \sum_{i_A,i_B=\uparrow,\downarrow} \left({\overline v}^{(i_A)}(p_A)\Gamma_I v^{(i_B)}(p_B)\right)^\dagger \left({\overline v}^{(i_A)}(p_A)\Gamma_{II} v^{(i_B)}(p_B)\right) \nonumber \\
& = {\rm Tr}\left[{\overline \Gamma}_I \left(\gamma^\mu {p_A}_\mu  - m_A\right)\Gamma_{II} \left(\gamma^\mu {p_B}_\mu - m_B\right)\right],
\end{align} for any two $4 \times 4$ matrices $\Gamma_I, \Gamma_{II}$, where
$\overline{\Gamma}_I \equiv \gamma^0 \Gamma_I^\dagger \gamma^0$, with $\gamma^\mu$ the $4\times 4$ Dirac $\gamma$-matrices.
For our purposes it suffices to calculate the width (\ref{width}) in flat Minkowski space time. In this case, we have the properties \begin{align}
	\{\gamma^\mu, \, \gamma^\nu \} &= 2\eta^{\mu\nu}\textbf{1}\\
	\{\gamma^\mu, \, \gamma^5 \} &= 0\\
	{\overline \gamma}^\mu \equiv \gamma^0 (\gamma^\mu)^\dagger \gamma^0 &= \gamma^\mu\\
	\overline{\gamma^\mu \gamma^5} \equiv \gamma^0 (\gamma^\mu \gamma^5)^\dagger \gamma^0 &= \gamma^\mu \gamma^5\\
	{\rm Tr}[{\rm \textbf{1}}] &= 4\\
	{\rm Tr}\left(\gamma^\mu \,\gamma^\nu \right) &= 4\eta^{\mu\nu}\\
	{\rm Tr}\left(\gamma^\mu \gamma^\lambda \gamma^\nu \gamma^\rho \right) &= 4\left(\eta^{\mu\lambda}\eta^{\nu\rho} - \eta^{\mu\nu}\eta^{\lambda\rho} + \eta^{\mu\rho}\eta^{\nu\lambda}\right)
\end{align} where $\gamma^5 = i \gamma^0 \, \gamma^1 \, \gamma^2 \, \gamma^3 $, the trace ${\rm Tr}$ is over spinor indices, \textbf{1} denotes the $4 \times 4$ identity  matrix in spinor space and $\eta^{\mu\nu} $ is the Minkowski metric with the signature convention $(\,+, \,-, \,-, \,-)$.
Note that the trace ${\rm Tr} \left( \gamma^\mu \gamma^\nu \gamma^5\right) = 0$ while ${\rm Tr} \left( \gamma^\mu\gamma^\lambda \gamma^\nu \gamma^\rho \gamma^5\right)$ is totally antisymmetric in the Lorentz indices.

Since the phenomenological considerations of \cite{amrr} require the vector boson mass $m_V$ to be much larger (at least four orders of magnitude)  than the sterile neutrino DM mass $m$,  $m_V \gg m $, we may treat the fermionic product of the decay $V_\mu \, \overline{\nu_{1}}\,  \nu_{1} $ as \emph{practically massless}. Hence, applying the identities of \cref{rules2} in this case,
yields: \begin{equation}
	- i {\cal M} =  \epsilon_W^\mu (p) \, \bar{u}_{f_1}^{(i1)} (p_1^\prime)  \, \left( -i g_V\, \gamma_\mu \, \right)  \, \left(\frac{1 + \gamma^5}{2}\right)   v_{f_2}^{(i2)} (p^\prime_2)\,,
\end{equation} where $\epsilon^\mu_V $ is the polarisation of the massive $V$-boson. and $f_{1,2}$ denote the fermionic decay products in the three processes. The fermions $f_i$ are all massless.
The square of the initial-spin-averaged and final-spin-summed amplitude entering \cref{width} reads:
 \begin{equation}\label{avampl}
   |\overline{{\cal M}}|^2 = \frac{1}{3} \sum_{i_1, i_2 = \uparrow, \downarrow} |{\cal M}|^2,
   \end{equation}
where the factor $1/3$ is due to the fact that we have $2 s + 1$ (with $s=1$)
initial spins of the massive vector boson to average over. Taking into account  the identities of \cref{rules2}, with the matrices $\Gamma_I, \Gamma_{II}$ being $\gamma^\mu, \gamma^\nu$ and $\gamma^\mu \gamma^5, \gamma^\nu \gamma^5$,
as well as the fact that the sum over vector-boson-$V$ polarisation (spin) states is \begin{equation*}
	\sum_{\rm spin} \epsilon_\mu (p) \, \epsilon_\nu (p) =  -\eta_{\mu\nu} + \frac{p_\mu p_\nu}{m_V^2} ,
\end{equation*} we may evaluate the amplitude (\ref{avampl}) as (from now on we omit the particle-species index from the polarisation tensors of spinors for simplicity)
\begin{align*}
	|\overline{{\cal M}}|^2 
	&= \frac{g_V^2}{3} \, \sum_{i_1, i_2 = \uparrow, \downarrow} \left[ \bar{u}^{(i1)} (p_1^\prime) \gamma_\mu \frac{1 +\gamma^5}{2} v^{(i2)} (p_2^\prime) \right]^\dagger \\
	&\phantom{=} \left[ \bar{u}^{(i1)} (p_1^\prime) \gamma_\nu \frac{1+ \gamma^5}{2} v^{(i2)} (p_2^\prime) \right]  \epsilon_V^\mu (p) \epsilon_V^\nu (p)\\
	&=\frac{g_V^2 }{3}  {\rm Tr} \, \Big[\gamma_\mu \frac{1 + \gamma^5}{2} (\gamma_\rho p_2^{\prime \, \rho}  ) \, \gamma_\nu \,
 \frac{1+ \gamma^5}{2} \, \gamma_\sigma \, p_1^{\prime \, \sigma} \Big]\\
	&\phantom{=} \times  \left( - \eta^{\mu \nu} + \frac{p_V^\mu p_V^\nu}{m_V^2} \right)\\
	&= \frac{g_V^2}{6} {\rm Tr} [\gamma_\mu (1+ \gamma^5) \, \gamma_\rho \gamma_\nu \gamma_\sigma  ] \ p_2^{\prime \, \rho} p_1^{\prime \, \sigma} \,  \left( - \eta^{\mu \nu} + \frac{p_V^\mu p_V^\nu}{m_V^2} \right)\, \\
	&= \frac{g_V^2}{6} {\rm Tr} [\gamma_\mu \, \gamma_\rho \gamma_\nu \gamma_\sigma  ] \ p_2^{\prime \, \rho} p_1^{\prime \, \sigma} \,  \left( - \eta^{\mu \nu} + \frac{p_V^\mu p_V^\nu}{m_V^2} \right)\, .
\end{align*} In the last simplification, we used the anti-commutation properties of $\gamma^5$ with $\gamma_\mu$, and the fact that $\left( \frac{1 + \gamma^5}{2} \right)^2 = \frac{1 + \gamma^5}{2} $ and $\frac{1+\gamma^5}{2} \frac{1-\gamma^5}{2} =0$.
Above we also took into account that the trace containing $\gamma^5$ is zero because it gives rise to a totally antisymmetric tensor (rubric) which is contracted with a symmetric tensor with respect to the $\mu, \nu$, indices,  $\left( - \eta^{\mu \nu} + \frac{p_V^\mu p_V^\nu}{m_V^2} \right)$.
Using the identities of Dirac matrices, given previously, we then obtain
\begin{align}\label{finalampl}
	|\overline{{\cal M}}|^2
	&= \frac{2\,g_V^2}{3} (p_{2 \mu}^\prime p_{1 \nu}^\prime + p_{2 \nu}^\prime  p_{1 \mu}^\prime  - \eta_{\mu \nu} p_1 ^\prime \cdot p_2^\prime  ) \left( - \eta^{\mu \nu} + \frac{p^\mu p^\nu}{m_V^2} \right) \nonumber\\
	&= \frac{2\,g_V^2}{3}\left( p_2^\prime \cdot p_1^\prime  + 2 \frac{p_2^\prime  \cdot p \, p_1^\prime \cdot p}{m_V^2} \right)
 \end{align}
where we used the on-shell condition for the $V$ momentum $ p^\mu \eta_{\mu\nu} p^\nu = m_V^2$.
Using the conservation of energy momentum in the vertex ($p$ incoming, $p_1^\prime, p_2^\prime$ outgoing),
\begin{equation}\label{cons}
 p^\mu - p_1^{\prime \, \mu} - p_2^{\prime \, \mu} = 0  ,
 \end{equation}
we square it (in a covariant way) to derive :
\begin{align*}
 0 	&=  p^\prime_2 \cdot p^\prime_2 + p \cdot p  +  p^\prime_1 \cdot p^\prime_1 + 2 p^\prime_2 \cdot p^\prime_1 - 2 p  \cdot (p^\prime_2 +  p^\prime_1) \\
		&= m_W^2  + 2 p^\prime_2 \cdot p^\prime_1 - 2 p \cdot p =  - m_W^2 + 2 p^\prime_2 \cdot p^\prime_1
\end{align*}\begin{equation*}
	\Rightarrow p^\prime_2 \cdot p^\prime_1 = m_W^2 /2
\end{equation*} where we used the on-shell conditions (in our conventions for the metric $(+1,-1,-1,-1)$) \begin{equation*}
	p \cdot p = m_V^2 , \quad p^\prime_1 \cdot p^\prime_1 = 0, \quad p^\prime_2 \cdot p^\prime_2 = 0 .
\end{equation*}
In a similar way by writing the square as \begin{align*}
	0	&=  p^\prime_2 \cdot p^\prime_2 + p \cdot p +  p^\prime_1 \cdot p^\prime_1 - 2 p_1^\prime  \cdot (p - p^\prime_2 )  - 2 p \cdot p^\prime_2 \\
		&= p^\prime_2 \cdot p^\prime_2 + p \cdot p  - p^\prime_1 \cdot p^\prime_1 - 2 p \cdot p^\prime_2 \\
		&\Rightarrow  p \cdot p^\prime_2 =  m_V^2 /2 .
\end{align*} And finally, by writing the square as: \begin{align*}
	0 &= p^\prime_2 \cdot p^\prime_2 + p \cdot p +  p^\prime_1 \cdot p^\prime_1 - 2  p^\prime_2 \cdot  (p - p^\prime_1 ) - 2 p^\prime_1 \cdot  p \\ 
		&= m_W^2 - 2 p^\prime_2 \cdot p^\prime_2  - 2 p \cdot p^\prime_1 \, \\
		&\Rightarrow \, p^\prime_1 \cdot p =  m_V^2 /2~.
\end{align*} Then, the  amplitude can be written in terms of masses
\begin{equation}
	\label{finalwidth}
	|\overline{{\cal M}}|^2  = \frac{2\,g_V^2}{3}\, m_V^2\, .
\end{equation}

In the rest frame of the $V$-boson ($E_V = m_V$,  $\vec{p}_V = 0$)
the  phase space integration in \cref{width} is done by first performing the spatial
delta function integration $$\int \d^3 p^\prime_{2}\, \delta ^{(3)} (  - p_1^\prime -  p_2^\prime )~, $$ which simply implies
that the spatial momenta of the decay products (which are massless particles) are equal in magnitude  $|\vec{p}_1^{\prime}| = |\vec{p}_2^{\prime}|$.

In the case of Majorana sterile neutrinos, there is one group of two ($n=2$) identical particles in the products of the $V$-vector-boson decay so the
statistical factor $\mathcal S$ in the definition of the width (\ref{width}) is
$\mathcal S=\frac{1}{2}$ (for Dirac type ``inos'' $\mathcal S=1$, as in that case there are no identical particles among the decay products). Treating the neutrino as practically massless inside the phase-space integration, which suffices for our approximate discussion here,
we then obtain: \begin{align}
	\Gamma 	&= \frac{1}{16 \, m_V} \int 4\pi |\vec p_2'|^2 \d|\vec p_2'|  \delta (M_V - 2|\vec{p_2'}|)\frac{1}{4\pi^2\, |\vec{p_2'}|^2} |\overline{{\cal M}}|^2 \nonumber \\
	\label{result}
					&= \frac{1}{32 \, \pi \, m_V} \, |\overline{{\cal M}}|^2 \simeq \frac{g_V^2} {48\, \pi}\, m_V
\end{align} where we used that $ \int \d|\vec p_2^\prime|   \delta (m_V - 2|\vec{p_2^\prime}|) = \frac{1}{2}$.

\bibliographystyle{JHEP} 
\bibliography{biblio}

\end{document}